\documentclass[notitlepage,aps,epsfig,showpacs,floats,onecolumn,amssymb,amsmath,floatfix,
groupedaddress,superscriptaddress]{revtex4-1}
\pdfoutput=1

\usepackage{graphicx}
\usepackage{dcolumn}
\usepackage{bm}
\usepackage{placeins}
\usepackage{float}
\usepackage{appendix}
\usepackage{bbold}
\usepackage{comment}
\usepackage[color=green!10, size=footnotesize]{todonotes}

\usepackage{graphicx} 
\usepackage{dcolumn} 
\usepackage{bm} 
\usepackage{float} 
\usepackage{bbm} 
\usepackage{color}
\usepackage{amsfonts}
\usepackage{ulem}
\usepackage{lmodern} 
\usepackage{amsmath} 
\usepackage{amsmath,amssymb}
\usepackage{natbib}
\usepackage{amsmath}
\newcommand\bi{\begin{itemize}}
\newcommand\ei{\end{itemize}}
\renewcommand{\vec}[1]{\boldsymbol{#1}}
\newcommand{\tens}[1]{\boldsymbol{#1}}
\newcommand{\bnabla}{\vec{\nabla}}

\usepackage[colorlinks=true,citecolor=blue]{hyperref}
\hypersetup{colorlinks=true,citecolor=blue,linkcolor=blue,urlcolor=black}
\begin{document}
\title{Stress-shape misalignment in confluent cell layers}

\author{Mehrana R. Nejad}
\affiliation{The Rudolf Peierls Centre for Theoretical Physics, Department of Physics, University of Oxford, Parks Road, Oxford OX1 3PU, United Kingdom}
\author{Liam J. Ruske}
\affiliation{The Rudolf Peierls Centre for Theoretical Physics, Department of Physics, University of Oxford, Parks Road, Oxford OX1 3PU, United Kingdom}
\author{Molly McCord}
\affiliation{Biophysics Program, University of Wisconsin--Madison}
\affiliation{Department of Mechanical Engineering, University of Wisconsin--Madison}
\author{Jun Zhang}
\affiliation{Biophysics Program, University of Wisconsin--Madison}
\affiliation{Department of Engineering Physics, University of Wisconsin--Madison}
\author{Guanming Zhang}
\affiliation{Center for Soft Matter Research, Department of Physics, New York University, New York 10003, USA}
\affiliation{Simons Center for Computational Physical Chemistry, Department of Chemistry, New York University, New York 10003, USA}
\author{Jacob Notbohm}
\affiliation{Biophysics Program, University of Wisconsin--Madison}
\affiliation{Department of Mechanical Engineering, University of Wisconsin--Madison}
\affiliation{Department of Engineering Physics, University of Wisconsin--Madison}
\author{Julia M. Yeomans}
\affiliation{The Rudolf Peierls Centre for Theoretical Physics, Department of Physics, University of Oxford, Parks Road, Oxford OX1 3PU, United Kingdom}

\maketitle

\noindent
{\bf Abstract}\\
~\\
This study investigates the relationship between cell shape and cell-generated stresses in confluent cell layers. Using simultaneous measurements of cell shape orientation and cell-generated contractile forces in MDCK and LP-9 colonies, we report the emergence of correlated, dynamic domains in which misalignment between the directors defined by cell shape and by contractile forces reaches up to 90$^o$, effectively creating extensile domains in a monolayer of contractile cells. To understand this misalignment, we develop a continuum model that decouples the orientation of cell-generated active forces from the orientation of the cell shapes. This challenges the prevailing understanding that cells throughout a tissue create either contractile or extensile forces, and the validity of the usual active nematic models of cell motility where active forces are strictly slaved to cell shape orientation.\\
~\\
~\\

\noindent
{\bf Main}\\
~\\
Cells are the fundamental building blocks of life, and their ability to collectively generate active forces plays a crucial role in physiological processes from morphogenesis \cite{sutlive2022generation}, tissue growth and repair \cite{martin2004parallels} to apoptosis \cite{toyama2008apoptotic}, tumour development \cite{jain2014role} and metastasis \cite{wirtz2011physics}. Confluent cell monolayers plated on adhesive substrates are widely used as model systems in investigations aimed at 
understanding collective cell motility \cite{trepat2009physical}. There is growing evidence that the dynamics of such confluent cell layers can often be well described by the theories of active nematics \cite{duclos2017topological,balasubramaniam2021investigating,saw2018biological,balasubramaniam2022active,PhysRevLett.129.098102}.

\begin{figure*}[!ht]
	\centering
	\includegraphics[width=14cm]{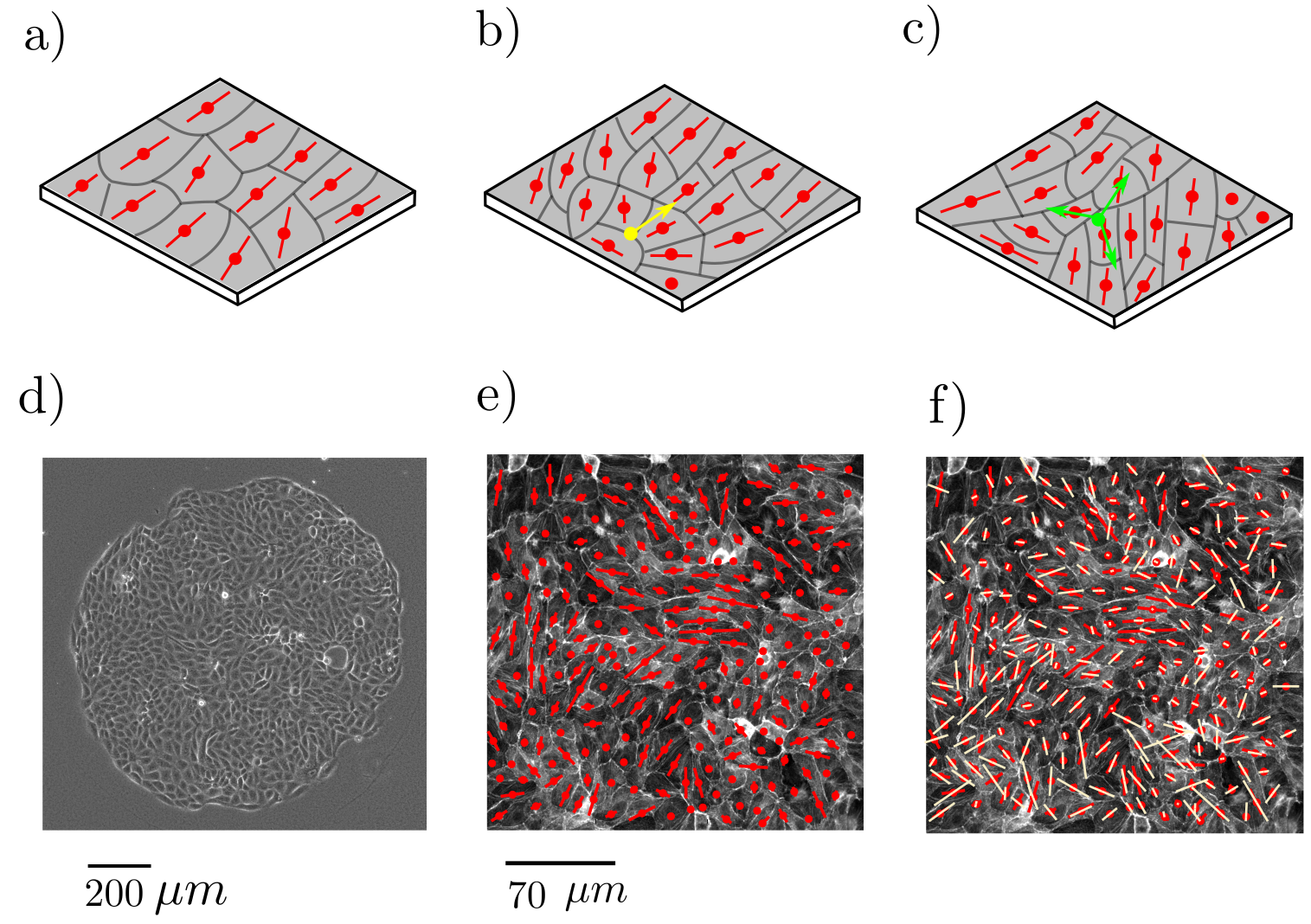}
	\caption{\footnotesize{(a) Nematic ordering of cell shape in a confluent cell monolayer. Red lines indicate the nematic directors $\mathbf{n}$, which lie along the long axis of a cell, and which tend to align parallel.
(b) Cell shape orientations around a $+1/2$ topological defect shown in yellow. 
(c) Cell shape orientations around a $-1/2$ topological defect shown in green.
(d) Geometry of the MDCK monolayers used in the experiments. 
(e) Example of a typical experiment measuring the long axis of cell shape orientation,  $\mathbf{n}$, (red lines) in an MDCK monolayer. 
(f) A typical set of results from 
experiments measuring the 
orientation of maximum contractile stress $\mathbf{m}$ (white lines) overlaid on the orientations of cell shape, $\mathbf{n}$ (red lines)}. In (e)-(f) the lengths of the red (white) lines are proportional to the deviation of the cell aspect ratio from unity (magnitude of the contractile stress).
}
\label{fig1}
\end{figure*}

Describing cells as {\em active} emphasises that they continuously take energy from their surroundings and use it to initiate life processes \cite{Ramaswamy_2017,RevModPhys.85.1143}. Nematic particles are elongated in shape, and {nematic ordering} occurs when their long axes tend to align parallel (Fig.~\ref{fig1}a). Such ordering has frequently been observed in long, thin cells such as fibroblasts \cite{duclos2017topological,li2017mechanism} or LP-9 \cite{zhang2021} but is more surprising in cell types that are on average isotropic, such as the Madin-Darby Canine Kidney (MDCK) cell line. Here extensions in cell shape, driven by active forces, are locally correlated to give nematic order \cite{comelles2021epithelial,saw2017topological,ascione2022collective}.

The combination of activity and nematic ordering leads to striking collective behaviours which are mirrored between active nematic models and cell monolayers. These include active turbulence, characterised by cell velocities that are chaotic with regions of high vorticity \cite{PhysRevLett.120.208101}, spontaneous directed flow in confinement \cite{duclos2018spontaneous}, and the identification of motile topological defects \cite{saw2017topological}, long-lived cell configurations at which domains of different cell orientations meet and the nematic order is frustrated. Fig.~\ref{fig1}b,c show the cell orientations around the +1/2 and -1/2  defects which predominate in 2D cell layers.

The direction of the long axis of a cell is an obvious way to define a preferred nematic shape axis. The local 
axis of principal stress gives an alternative way to choose the local axis of any nematic ordering. In almost all work to date it has been assumed that the two definitions are equivalent, meaning that the stress and shape axes are tightly coupled \cite{Simha02,Doostmohammadi18,comelles2021epithelial,PhysRevX.12.041017}. Under this assumption, differences between the axes of stress and shape would be modest and attributable to biological noise.
Here we challenge this assumption and show that there are dynamic, correlated regions in cell layers,  where the stress and shape axes are {systematically} misaligned. Introducing the possibility of such misalignment in a continuum model of active nematics reproduces the temporal and spatial correlations and misalignment angle observed in  
our experiments and emphasises the key role of active flows in driving the misalignment. \\

Our experiments were performed on confluent MDCK layers of diameter 1 mm plated on polyacrylamide substrates with a Young's modulus of 6 kPa (Fig.~\ref{fig1}d). We define the cell shape orientation in the tissue by assigning each cell a director (headless vector) $\mathbf{n}$  which lies along the long axis of the anisotropic cell shape as shown in Fig.~\ref{fig1}e \cite{DeGennesBook}. 
Monolayer Stress Microscopy was used to measure the stress tensor $\bm{\sigma}$, from which we computed the orientation of the first principal stress, which defines the local orientation $\mathbf{m}$ along which contractile forces are generated (Fig.~\ref{fig1}f).  By interpolating between individual cells we obtain continuous director fields $\mathbf{n}$ and $\mathbf{m}$ which respectively describe cell shape orientation and the principal axis of contractile stress throughout the tissue (see Methods).

\begin{figure*}[!ht]
	\centering
	\includegraphics[width=18cm]{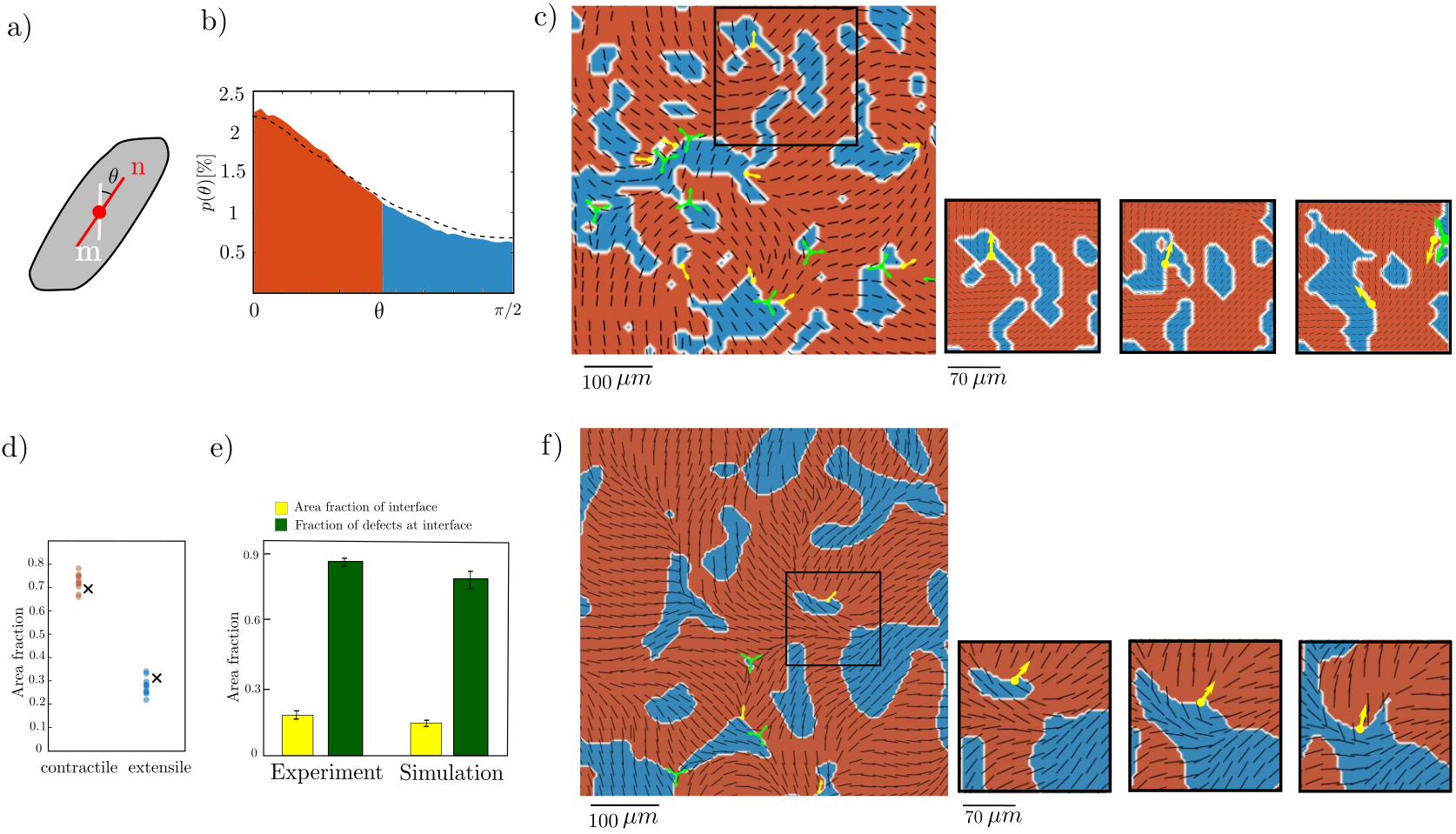}
	\caption{\footnotesize{(a)  Definition of the misalignment angle $\theta$ between the shape orientation axis $\mathbf{n}$, and the principal axis of contractile stress $\mathbf{m}$. 
 (b) Distribution of the misalignment angle $\theta$. Red/blue colouring denotes contractile ($\theta<45^o$)/extensile $ (\theta>45^o$) values. The distribution contains data from
 11 independent MDCK islands and all time points. The black dotted line shows the result from simulation.
 (c) left: Tissue snapshot with the cell orientation field $\mathbf{n}$ shown as black lines on top of a colour map distinguishing contractile (red) and extensile (blue) regions. Topological defects in the cell orientation are indicated by yellow ($+1/2$) and green ($-1/2$) symbols.  Right: Snapshots of a region of the same tissue taken 15 minutes apart showing the evolution of extensile clusters.  The time axis is from left to right.
	(d) Experimental time average of the area fraction of contractile (red) and extensile (blue) cells for 11 different cell islands. The crosses show the results from simulations. 
 (e) Defects are preferentially found in the vicinity of boundaries between extensile and contractile regions. 
 (f) Snapshot from simulations with the cell orientation field $\mathbf{n}$ shown as black lines on top of a colour map distinguishing contractile (red) and extensile (blue) regions. Topological defects in the cell orientation are indicated by yellow arrow ($+1/2$) and green trefoil ($-1/2$) symbols. The time axis is from left to right. } }
	\label{fig2}
\end{figure*}

We analysed the local cell shape and stress orientations over the course of 12 hours in 11 MDCK tissue samples, in time lapse experiments taking data every 15 minutes. 
We define $\theta$ as the misalignment angle between the local cell shape orientation $\mathbf{n}$ and the principal axis of contractile stress $\mathbf{m}$ in the tissue (Fig.~\ref{fig2}a). The distribution of $\theta$ is shown in Fig.~\ref{fig2}b. While most cells create contractile stresses along their cell shape axis ($\theta \approx 0$), there is a large number of cells in which the axis of contractile stress is significantly misaligned with respect to shape orientation. If the misalignment angle reaches $\theta \approx 90^\circ$, cells create contractile stresses perpendicular to the cell shape orientation, thereby pulling inward not along their long shape axis but rather along their short shape axis. 
In the following we will refer to cells with large misalignment ($\theta > 45^\circ$) as \textit{extensile} and cells with small misalignment ($\theta < 45^\circ$) as \textit{contractile} following the usual terminology in the mathematics and active matter literature, and we distinguish these ranges of $\theta$ as blue and red in Fig.~\ref{fig2}b. 

We now investigate the spatial and temporal correlation of the shape and the stress orientations in the MDCK monolayers.  Fig.~\ref{fig2}c (left) shows a tissue snapshot 
where the cell shape orientation field $\mathbf{n}$ is shown as black lines and the color map again indicates whether  $\theta$ is greater (extensile, blue) or less (contractile, red) than $45^\circ$. Fig.~\ref{fig2}c (right) shows the time evolution of a region of the tissue with snapshots taken at 15-minute intervals. Similar data is presented dynamically in Movie 1.
It is evident that the misalignment angle 
forms evolving spatio-temporal patterns 
where extensile cells form small, dynamic clusters in a mostly contractile background. The extensile clusters grow, shrink and coalesce over time. 
 The time-averaged area fraction of extensile cells is $27 \pm 4\%$ (Fig.~\ref{fig2}d).

\begin{figure*}[!ht]
	\centering
	\includegraphics[width=17cm]{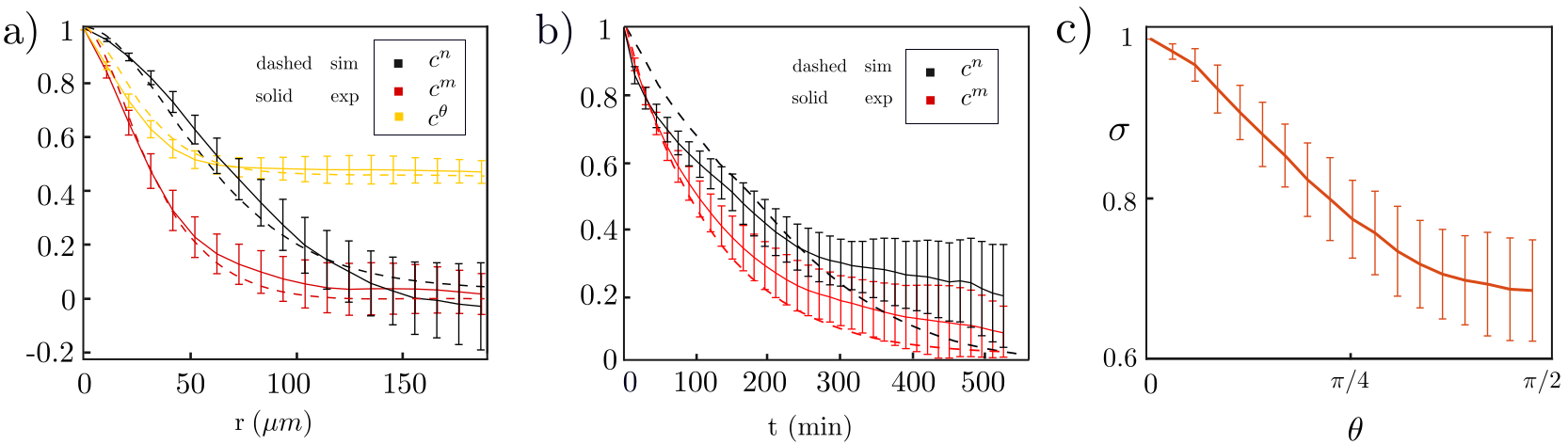}
	\caption{\footnotesize{(a) Decay of the spatial correlation functions, $C^{x}(r)=\langle \cos 2[\psi_x(r+r_0,t_0)- \psi_x(r_0,t_0)] \rangle_{t_0,r_0}$,   where $\psi_x$ represents shape director angle $C^{\bf{n}}$, stress angle $C^{\bf{m}}$, or the mismatch angle $C^{\theta}$. 
  Comparison of the simulation and the experiment assumes $10 LB$ spatial units $\sim 44 \mu m$. (b)  Decay of the time correlation functions, $C^{x}(t)= \langle\cos 2[\psi_x(r_0,t+t_0)- \psi_x(r_0,t_0)] \rangle_{t_0,r_0}$,
 where $\psi$ represents shape director angle $C^{\bf{n}}$ or the stress angle $C^{\bf{m}}$.
	 Comparison of the simulation and the experiment assumes $100 LB$ time units $\sim 10 \: \mbox{min} $. (c) Magnitude of the stress $\sigma$ as a function of the misalignment angle $\theta$. The magnitude of the stress is re-scaled with its maximum value in each experiment before averaging. In all plots the error bars show the standard deviation over 11 experiments.
 }}
	\label{fig3new}
\end{figure*}

 To further quantify the spatial patterns we calculated the spatial and time correlation 
 functions of the cell shape orientation, the cell stress orientation, and the mismatch angle $\theta$. These are defined as
 \begin{equation}
C^{x}(r)=\langle \cos 2[\psi_x(r+r_0,t_0)- \psi_x(r_0,t_0)] \rangle_{t_0,r_0},\;\;\;\;\;\;\;\; C^{x}(t)= \langle\cos 2[\psi_x(r_0,t+t_0)- \psi_x(r_0,t_0)] \rangle_{t_0,r_0}, 
 \label{corr}
 \end{equation}
 where $\psi_x$ represents the shape director angle, stress angle, or the mismatch angle $\theta$, and
  $\langle \ldots \rangle_{r_0,t_0}$ denotes an average over space (a circle of diameter $ 312 \mu m$ in the centre of the island to avoid edge effects) and time. 
  The spatial correlations, shown in Fig.~\ref{fig3new}a, indicate a length-scale $\sim 50 \mu m$ for the cell stress orientation, and a longer length scale $\sim 100 \mu m$ for the cell shape orientation. From the time correlation functions, in Fig.~\ref{fig3new}b, we identify a time-scale for the decay of 
 the extensile patches $\sim 300$ minutes.\\ 
 
Many cells contain bundles of actomyosin, termed stress fibres, that tend to form along the long axes of cells and are the primary source of contractile stresses \cite{Pellegrin07}.  We hypothesise that the regions of large misalignment angle are due to active flows disturbing the natural alignment of the stress fibres with the long axis of the cell due to different responses of the shape axis $\mathbf{n}$ and the stress axes $\mathbf{m}$ to flows.  This leads to the formation of 
extensile regions which have a large mismatch between the shape and stress axes. A cell's stress axis and shape axis then gradually relax towards each other.

Continuum tissue models, based on the equation of motion of active nematics, have been very successful in explaining cell motility on a coarse-grained level \cite{balasubramaniam2022active,balasubramaniam2021investigating,duclos2017topological,saw2018biological,saw2017topological}. However, the assumption has always been that the principal axis of contractile stress $\mathbf{m}$ and the shape axis $\mathbf{n}$ are indistinguishable. Therefore, to investigate the consequences of our hypothesis, we extend the continuum modelling to decouple $\mathbf{m}$ and $\mathbf{n}$. 

We describe the shape of the cells and the stress using nematic order parameters $\mathbf{Q}^n = S^n (\mathbf{n} \otimes \mathbf{n}-\mathbf{I}/2)$ and
$\mathbf{Q}^m = S^m (\mathbf{m} \otimes \mathbf{m}-\mathbf{I}/2)$, respectively \cite{DeGennesBook}.  The nematic order parameters encode the magnitude of nematic order in cell shape $S^n$ or in the stress  $S^m$, and the director field associated with cell shape,  $\mathbf{n}$ or the stress $\mathbf{m}$. We assume that the flows are created by contractile active forces that act along the direction of stress fibres $\mathbf{m}$. The active flows advect the cells, and the vorticity of the flows rotates both the shape and the stress director fields.
The experimental spatial correlation functions show that the nematic order of the shape director $\mathbf{n}$ has a longer length scale than that of the stress director $\mathbf{m}$ which we model by choosing different elastic constants in the free energy. This also means that the shape and the stress directors respond in different ways to the vortical flows leading to misalignment between $\mathbf{m}$ and $\mathbf{n}$. We include a term in the free 
energy which acts to slowly relax the misalignment.
In the continuum simulations, we have the freedom to choose a length and time scale, and we do this by matching the length and time scales of the correlation functions (Eq.~\ref{corr}) between simulation and experiment, as shown in 
Figs.~\ref{fig3new}a and~\ref{fig3new}b. See Methods for further details of the modelling.
\begin{figure*}[!ht]
	\centering
	\includegraphics[width=18.3cm]{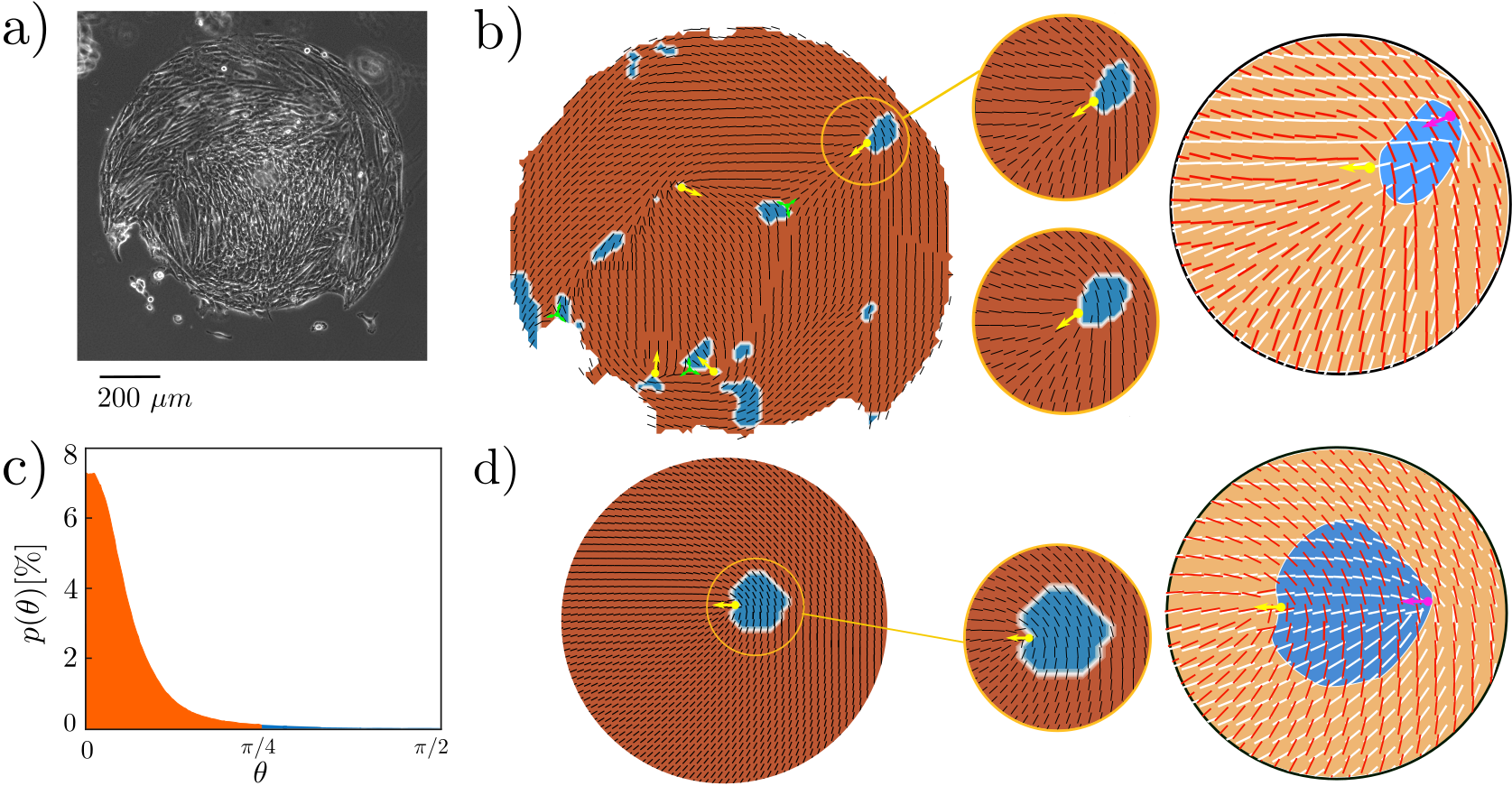}
	\caption{\footnotesize{(a) Snapshot of an LP-9 island. (b) left: Tissue snapshot with the cell shape orientation field $\mathbf{n}$ shown as black lines on top of a colour map distinguishing contractile (red) and extensile (blue) regions. Topological defects in the cell shape orientation are indicated by yellow ($+1/2$) and green ($-1/2$) symbols.  Middle: Snapshots of a region of the same tissue taken 40 minutes apart showing that the extensile cluster remains fixed. The time axis is from top to bottom.  Right: Close-up view of the same tissue. Cell shape orientation is shown in red and stress orientation in white. The colormap shows the extensile (blue) and contractile (orange) regions (smoothed). The defect in the cell shape (stress) orientation is shown in yellow (magenta). (c) Distribution of the misalignment angle $\theta$. Red/blue colouring denotes contractile ($\theta<45^o$)/extensile $ (\theta>45^o$) values. The LP-9 cells predominantly form contractile regions (compare with Fig.~2b for MDCK cells). (d) Left: Snapshot from simulations with the cell shape orientation $\mathbf{n}$ shown as black lines on top of a colour map distinguishing contractile (red) and extensile (blue) regions. The +1/2 defect in the cell shape is indicated by a yellow arrow. Right: Close-up view of the same tissue. Cell shape orientation is shown in red and stress orientation is white. The colormap shows the extensible (blue) and contractile (orange) regions (smoothed). The defect in the cell shape (stress) orientation is shown in yellow (magenta). }}
	\label{lp9}
\end{figure*}

We compare the simulation results to the experiments in Fig.~\ref{fig2}. In agreement with the experiments, spatially correlated domains of extensile cells in a contractile background emerge in the simulations (Fig.~\ref{fig2}f and Movie 2). We also obtain a quantitative match to the probability distribution for the misalignment angle $\theta$ if we
use a realignment time scale of 25 minutes (Fig.~\ref{fig2}b).

Active turbulence is also characterised by topological defects and we noticed that topological defects in the cell orientation field tend to sit at the interfaces between extensile and contractile domains. This is illustrated in Figs.~\ref{fig2}c (experiment) and ~\ref{fig2}f (simulations) where $+1/2$ defects are indicated by a yellow arrow and $-1/2$ defects by a green trefoil. To quantify the results, we consider the points that are
 at a distance smaller than $r<5.2 \mu m$ (equal to the grid size after interpolation) from the interface as the \textit{interface domain} while the rest of the tissue is defined as the \textit{bulk domain}. A spatial and temporal average shows that interface domains make up about $18\pm 2\%$ of the tissue area fraction in the experiments, and about $15\pm 3\%$  of the area fraction in simulations. However, about $86 \pm 2 \%$ ($86\pm 5 \%$) of defects lie within interface domains in experiments (simulations). These results are shown in Fig.~\ref{fig2}e.

Our interpretation
leads to the prediction that in the extensile regions the stress fibres will be in the process of re-forming to realign with the new direction of cell elongation, and are therefore less efficient in producing contraction. In Fig.~\ref{fig3new}c we plot the magnitude of the contractile stress as a function of $\theta$ showing a clear decrease. 
Moreover, to check our interpretation visually we fixed cells for fluorescent imaging of actin fibres in a selection of samples. We found that it was rarely possible to visually ascertain an unambiguous direction of the stress fibres in cells with large $\theta$ whereas stress fibres were much clearer in cases when they were aligned along the long axis of the cell (see Supplementary Material). \\

We next, as a comparison, performed similar experiments on the human mesothelial cell line LP-9 (Fig.~\ref{lp9}a).  These cells, which have an elongated morphology with a high aspect ratio, showed a behaviour which contrasted with the MDCK islands. A very small number of topological defects were present at the beginning of the experiments. These persisted and remained in approximately the same position throughout the experiments (40 hours) and no new defects were created, indicating that the cell layer was behaving primarily as a passive nematic, with any active flows not sufficiently strong to create defects or substantially change the cell orientation.

There was, however, still a population of extensile cells with $\theta>45^\circ$, but 
it was far smaller than in the MDCK monolayers. The extensile cells formed small ($3.4 \%$ of the area of the tissue) clusters (Fig.~\ref{lp9}b) adjacent to the defects in the cell shape director field. We found no evidence that the extensile regions disappear within the time scale of the experiment. 
 
As shape director $\bf{n}$ does not change we modelled the LP-9 cells by fixing a defect in cell shape $\bf{n}$ at the center of a circular cellular island and 
allowed the stress field $\bf{m}$ to relax. An extensile region 
was indeed formed 
next to the defect in cell shape $\bf{n}$, as shown in Fig.~\ref{lp9}c. This dynamical steady state is a result of the balance between the
 elastic energy, which favours
 nematic alignment of the stress directors $\bf{m}$, and the term which encourages $\bf{m}$ to align with $\bf{n}$. 
The exact position and size of the extensile region relative to the defect varies, depending on the initial condition for $\bf{m}$, indicating the existence of metastable solutions.\\

We have shown that active flows and the different elastic properties of cell shapes and active stress filaments lead to long-lived regions of large shape-stress misalignment within cell monolayers. Experimental results on MDCK and LP-9 cells are interpreted in terms of an active nematic model which corrects 
the usual assumption that the shape and stress axes are identical, but rather allows them to respond differently to flows causing misalignment between the shape and stress axes. In MDCK monolayers the shape and stress slowly realign, thus introducing a time-scale that must be taken into account when describing the monolayer dynamics.
Our results question the common practice of using the direction of defect velocities to identify cell monolayers as contractile or extensile. 

\newpage
\noindent
{\bf Methods}\\

\noindent
{\bf Cell culture:}
Madin-Darby canine kidney (MDCK) type II cells transfected with green fluorescent protein (GFP) attached to a nuclear localization signal (a gift from Professor David Weitz, Harvard University) were maintained in low-glucose Dulbecco's modified Eagle's medium (10-014-CV, Corning Inc., Corning, NY) with 10\% fetal bovine serum (FBS, Corning) and 1\% G418 (Corning).  LP-9 cells (AG07086) were maintained in a 1:1 ratio of Medium 199 and Ham F-12 (Corning) with 10 ng/ml epithelial growth factor (MilliporeSigma) and 0.4 $\mu$g/ml hydrocortisone (MilliporeSigma) that was supplemented with 15\% FBS. All cells were maintained in an incubator at 37$^\circ$C and 5\% CO$_2$.\\
~\\
{\bf Time lapse imaging and analysis:}
Polyacrylamide gels embedded with fluorescent particles (580/605, diameter 0.5 $\mu$m, carboxylate modified; Life Technologies) were fabricated with Young's moduli of 6 kPa and thickness of 150 $\mu$m using methods described previously \cite{saraswathibhatla2020, saraswathibhatla2020scidata}. 
Polydimethysiloxane (PDMS, Sylgard 184) was cured in 200 $\mu$m thick sheets. The sheets were cut into 20$\times$10 mm squares and then 1 mm holes were cut using a 1 mm biopsy punch. The PDMS masks were adhered to the gels using previous methods \cite{saraswathibhatla2020}, and the 1 mm circular hole was coated with 0.01 mg/ml type I rat tail collagen I (BD Biosciences) with the covalent crosslinker sulfo-SANPAH (Pierce Biotechnology).  MDCK and LP-9 cells were seeded onto 1 mm islands 24 hr before imaging. The cells and particles were imaged using an Eclipse Ti-E microscope (Nikon, Melville, NY) with a 10$\times$ numerical aperture 0.5 objective (Nikon) and an Orca Flash 4.0 digital camera (Hamamatsu, Bridgewater, NJ) running on Elements Ar software (Nikon). All imaging was done at 37$^\circ$C and 5\% CO$_2$. The cells were imaged every 15 or 20 min for 24 hr. After imaging, the cells were removed by incubating in 0.05\% trypsin for 1 hr, and images of the fluorescent particles in the substrate were captured for a traction-free reference state for traction force microscopy \cite{dembo1999}. To this end, Fast Iterative Digital Image Correlation was used\cite{Bar-Kochba2015} followed by Fourier transform traction microscopy \cite{butler2002} accounting for finite substrate thickness \cite{delalamo2007,trepat2009}. The displacements were computed using 32$\times$32 pixel subsets (21$\times$21 $\mu$m$^2$) with a spacing of 8 pixels (5.2 $\mu$m). Stresses within the monolayer were computed with monolayer stress microscopy \cite{tambe2011, tambe2013}. Cell orientations were determined using the ImageJ plugin OrientationJ.\\
~\\
{\bf Imaging and analysis for fixed cells:}
\begin{figure*}[b] 
    \centering
     \includegraphics[width=0.6\textwidth]{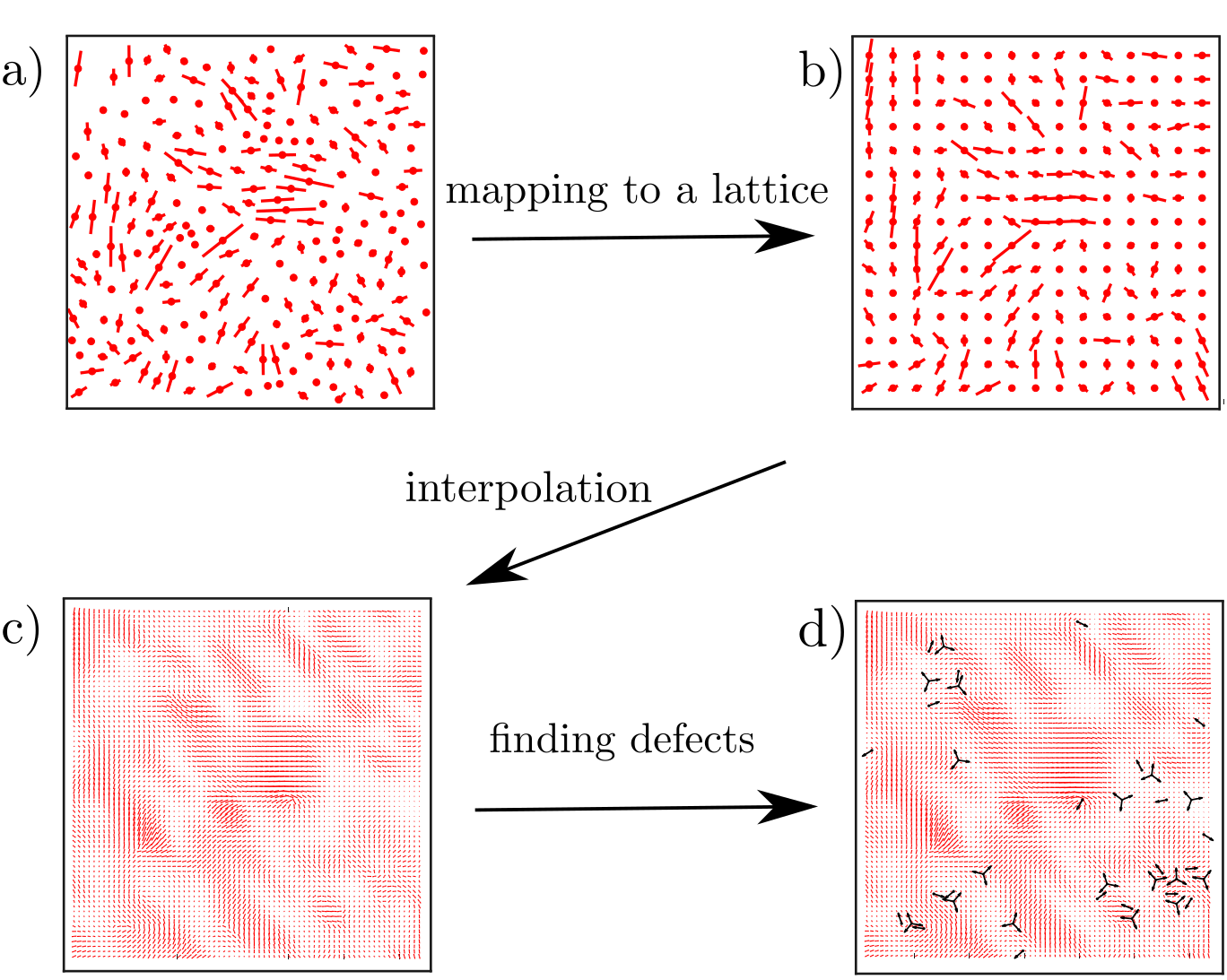}
    \caption{\footnotesize{Construction of the director fields for cell orientation and contractile stress: a) Cell position and orientation, as shown in Fig. \ref{fig1}. b) Cell position and orientation after mapping to a lattice. The lattice unit is equal to $dl= L/\sqrt{N}$, where $L$ is system size and $N$ is the number of cells in the snapshot. c)  Linear interpolation is used to find the director field on a lattice with a smaller mesh. This step is required to find the position and orientation of defects correctly. d) Using the director on the new lattice, we then use the defect-identifying algorithm introduced in Ref. \cite{vromans2016orientational} to find the direction and orientation of $\pm 1/2$ defects. }}\label{f3}
\end{figure*}
Imaging actin stress fibers required fixing the cell monolayers, which had to be done after collecting data for traction and stress measurements \cite{mertz2012, saraswathibhatla2021}. To this end, reference images of the fluorescent particles were collected before seeding the cells. Polyacrylamide gels were made as described above, and reference images of the fluorescent particles in the reference (undeformed) state were collected. Microscopy was the same as described above with the exception of using a 20$\times$ numerical aperture 0.5 objective (Nikon). MDCK cells were seeded and allowed to come to confluence overnight. The cell culture medium was changed 1 hr prior to imaging, and cells were imaged every 10 min for 1 hr. Cell monolayers were fixed using chilled 4\% paraformaldehyde solution for 20 min. Cells were then stained for actin using ActinRed 555 ReadyProbe Reagent (Invitrogen, catalog number R37112) according to manufacturer instructions and images. To analyse the orientation of the fixed cells, the cells were manually segmented based on images of the cell cortex. The orientation of actin stress fibers was identified manually for cells having visually clear stress fibers. Stresses were computed using monolayer stress microscopy, as described above, but for these experiments, the stresses at the boundaries were unknown, meaning that the recovered stresses represented not the full stress tensor but rather deviations from the average of the stress tensor at the boundaries. Our prior experiments with this cell type show that the average stress tensor is nearly isotropic (indicating small shear stresses) \cite{notbohm2016, saraswathibhatla2020scidata}. If the stress state at the boundary is isotropic, then there is no error in computing the orientation of principal stresses and the traceless stress tensor, both of which are reported in the main text for experiments using fixed imaging.\\
~\\
{\bf Construction of the director fields for cell orientation and contractile stress:} 
Fig.~\ref{f3} summarise the process of defining smooth director fields and identifying the position and orientation of defects. First, cell positions and orientations are mapped onto a square lattice with a lattice constant $dl = L/\sqrt{N}$, where $L$ and $N$ are the system length and number of the cells in the snapshot, respectively. To find the orientation of the cells on a lattice, each cell is mapped to the closest lattice site. The result is shown in Fig.~\ref{f3}(b). In the original lattice built by the cells, the lattice spacing is large and that makes it impossible to find the position and orientation of the defects accurately. As a result, we need to find the cell orientation on a lattice with a smaller mesh size. We construct this lattice by linear interpolation between lattice points (Fig.~\ref{f3}c). Using the interpolated director field, we can then use a defect finding algorithm to find the position and orientation of $\pm 1/2$ defects (Fig.~\ref{f3}d).

The stress matrix
measured in experiments has three independent components $\sigma_{xx}$, $\sigma_{xy}$ and $\sigma_{yy}$. It has a non-zero trace, and we first make it traceless by adding a constant $c$ to the diagonal elements, so that $\sigma_{yy}+\sigma_{xx} + 2 c =0$. To find the orientation of the contractile stress, we find the  two mutually perpendicular axes  which are
parallel to the 
orientations of positive and negative principal stress. These axes can be found by a rotation of the stress matrix through an angle $\theta_p$ such that the shear stress $\sigma_{xy}$ becomes zero. We note that there are two directions over which the shear stress becomes zero, 
$\theta_p$ and $\theta_p + \pi/2$. We define the orientation of contractile cell-generated stress to be along the orientation of the positive stress (pointing outwards). \\
~\\
{\bf The continuum model for active cell monolayers:} 
We describe the motion of cells within the monolayer by a continuous velocity field $\mathbf{u}$. The local orientation of cell shapes is described by a tensor field $\textbf{Q}^n= S^n (\mathbf{n} \mathbf{n} -\textbf{I}/2)$ and the orientation of the stress fibres by a second tensor field $\textbf{Q}^m= S^m (\mathbf{m} \mathbf{m} -\textbf{I}/2)$. These nematic order parameters encode the magnitude of nematic order in the cell shape, $S^n$, or the magnitude of the contractile stress, $S^m$, and the director field associated with the cell shape,  $\mathbf{n}$, or the stress, $\mathbf{m}$ \cite{DeGennesBook}. This description differs from previous active nematic continuum theories in that it allows for a finite misalignment angle $\theta = \cos^{-1}(\mathbf{n} \cdot \mathbf{m})$ between the elongation axis of cell shape and the axis along which contractile forces are generated by stress fibres (see Fig.~2a).

Following empirical arguments, we assume that in equilibrium the cell monolayer is governed by the following free energy density:
\begin{equation}
  f = \frac{C}{2} (1-3 \tens{Q}^n : \tens{Q}^n)^2 + \frac{C}{2} (1-3 \tens{Q}^m : \tens{Q}^m)^2 + \frac{K_n}{2} |\bnabla \tens{Q}^n|^2  + \frac{K_m}{2} |\bnabla \tens{Q}^m|^2+ \frac{J}{2} (1-3 \tens{Q}^n : \tens{Q}^m)^2.
    \label{eq:enegry}
\end{equation}
In the absence of activity, the first and the second terms lead to a phase with nematic order in cell shape and cell stress. The third and the fourth terms penalise gradients in the shape and stress orientations, respectively. These are motivated by the observations of nematic ordering in both $\bf{n}$ and $\bf{m}$ in the experiments.
The final term tends to align shape and stress orientations. 

The shape orientation, $\bf{n}$, and the stress orientation, $\bf{m}$, change in response to active flows. Since they have different elastic constants, they respond differently to the active flows which can lead to a mismatch between their directions.
The dynamics of the nematic tensors is governed by ~\cite{beris1994thermodynamics}
\begin{align}
\left(\partial_t + \vec{u}\cdot\bnabla \right) \tens{Q}^n  &= -\tens{\Omega}^{'}_n\cdot \tens{{Q}}^n +\tens{{Q}}^n \cdot \tens{\Omega}^{'}_n + \gamma \tens{H}^n, \label{qnqe}\\
\left(\partial_t + \vec{u}\cdot\bnabla \right) \tens{Q}^m  &= -\tens{\Omega}^{'}_m\cdot \tens{{Q}}^m +\tens{{Q}}^m \cdot \tens{\Omega}^{'}_m+\gamma \tens{H}^m, \label{qmqe}
\end{align}
where $\gamma$ is the rotational diffusivity, $\tens{\Omega}^{'}_m=\tens{\Omega}+\Delta_m$, $\tens{\Omega}^{'}_n=\tens{\Omega}+\Delta_n$, $\tens{\Omega}=(\bnabla\vec{u}^{T}-\bnabla\vec{u})/2$ is the fluid vorticity, $\Delta_{m/n}$ is a uniform noise in the rotation, the molecular field $\tens{H}^x = -(\frac{\delta f}{\delta\tens{Q}^x} - \frac{\tens{I}}{2} \; \text{Tr}\frac{\delta f}{\delta\tens{Q}^x})$ shows the relaxation of the orientational order to the minimum of the free energy, and we have set the flow tumbling parameter equal to zero.

We assume that the flows observed in confluent cell monolayers can be well approximated by 
\begin{equation}
	\rho \left( \partial_{t} + \mathbf{u} \cdot \bnabla \right) \mathbf{u} = \bnabla \cdot \mathbf{\Pi} , \:
	\label{eq:NSE}
\end{equation}
where $\rho$ is mass density and the stress tensor  $\mathbf{\Pi} = \mathbf{\Pi}_{passive} + \mathbf{\Pi}_{active}$ includes a passive and an active contribution, where the passive, viscous terms $\mathbf{\Pi}_{passive}$ are well known from liquid crystal hydrodynamics \cite{beris1994thermodynamics,Marenduzzo2007}. Flows in confluent cell layers are predominantly driven by active dipolar forces created on the single-cell level by stress fibers which convert chemical energy into mechanical work. 
This gives an active term in the stress tensor,
\begin{equation}
	\Pi_{act} = - \zeta \mathbf{Q}^m \: .
	\label{eq:activestress}
\end{equation}
where we choose $\zeta<0$ to correspond to contractile forces.

The equations are solved using a hybrid lattice Boltzmann algorithm \cite{Marenduzzo2007}. The MDCK simulations are performed in a $200\times200$ box with periodic boundary conditions over $120000$ lattice-Boltzmann time-steps, and data is collected every 300 time-steps. The measurements are performed in steady state when the mean number of defects and the fraction of the extensile area do not, change over time. The initial orientation of both  $\bf{n}$ and $\bf{m}$ is random, and the magnitude of the order is  $S^n=S^m=1$. We choose values of parameters that lead to an active fluid in a low Reynolds number regime: $\rho=40, \eta= 20/3$. Other parameter values are: $\gamma=0.4, K_m=0.005, C=10^{-3}/3, \zeta=-0.03, K_n=0.065,J=0.008,$ and $\Delta_{n}$ and $\Delta_{m}$ are uniform random numbers in the interval $[-0.001 ,0.001]$.

We set up the LP-9 simulations such that nematic order only forms inside a circular region with radius $R=80$. The free energy in this region is again given by Eq.~(\ref{eq:enegry}), but the bulk free energy outside the circle is 
\begin{equation}
  f_{bulk} = \frac{C^{'}}{2} (\tens{Q}^n : \tens{Q}^n)^2 + \frac{C^{'}}{2} (\tens{Q}^m : \tens{Q}^m)^2
    \label{eq:enegrynew}
\end{equation}
which sets the magnitude of the order to be zero. We impose a defect in the shape director $\bf n$  at the centre of the inner region by setting the director angle with the $x$-axis to be equal to $\phi/2$, where $\phi$ is the polar angle in the co-ordinate system centered at the defect core. We do not allow $\bf{n}$ to vary in time; $\bf{m}$ relaxes towards the minimum of the free energy. We use the same parameter values as for the MDCK cells except $K_m=0.02, K_n=0.01,\zeta=0,\Delta_m=\Delta_n=0, C^{'}=0.003$.

We note that the quantitative fits achieved using this model must be viewed with some caution as there are several adjustable parameters. However, the qualitative behaviour is insensitive to the numerical values of the parameters.
\\
~\\

\noindent
{\bf Movie captions:}
\\
~\\
Movie 1: Tissue dynamics with the cell orientation field $\mathbf{n}$ shown as black lines on top of a colour map distinguishing contractile (red) and extensile (blue) regions. Topological defects in the cell orientation are indicated by yellow ($+1/2$) and green ($-1/2$) symbols.
\\
~\\
Movie 2: Dynamics in the simulations, with the cell orientation field $\mathbf{n}$ shown as black lines on top of a colour map distinguishing contractile (red) and extensile (blue) regions. Topological defects in the cell orientation are indicated by yellow ($+1/2$) and green ($-1/2$) symbols.\\
~\\

\noindent
{\bf Contributions:}
MRN, LR, GZ and JMY formulated the model and interpreted the results. MRN, MM, JZ, and JN analysed the experimental data. MRN performed the simulations. JN designed and supervised the experiments. MM and JZ performed the experiments.  LR, MRN, and JMY drafted the manuscript. All authors commented on the manuscript.
\\
~\\
\noindent
{\bf Acknowledgements:}
M. R. Nejad acknowledges the support of the Clarendon fund scholarship. L. J. Ruske acknowledges the support of the European Commission’s Horizon 2020 research and innovation programme under the Marie Sklodowska-Curie grant agreement No 812780. This project was funded by the National Science Foundation grant number CMMI-2205141.

\bibliography{references}

\begin{thebibliography}{40}%
\makeatletter
\providecommand \@ifxundefined [1]{%
 \@ifx{#1\undefined}
}%
\providecommand \@ifnum [1]{%
 \ifnum #1\expandafter \@firstoftwo
 \else \expandafter \@secondoftwo
 \fi
}%
\providecommand \@ifx [1]{%
 \ifx #1\expandafter \@firstoftwo
 \else \expandafter \@secondoftwo
 \fi
}%
\providecommand \natexlab [1]{#1}%
\providecommand \enquote  [1]{``#1''}%
\providecommand \bibnamefont  [1]{#1}%
\providecommand \bibfnamefont [1]{#1}%
\providecommand \citenamefont [1]{#1}%
\providecommand \href@noop [0]{\@secondoftwo}%
\providecommand \href [0]{\begingroup \@sanitize@url \@href}%
\providecommand \@href[1]{\@@startlink{#1}\@@href}%
\providecommand \@@href[1]{\endgroup#1\@@endlink}%
\providecommand \@sanitize@url [0]{\catcode `\\12\catcode `\$12\catcode `\&12\catcode `\#12\catcode `\^12\catcode `\_12\catcode `\%12\relax}%
\providecommand \@@startlink[1]{}%
\providecommand \@@endlink[0]{}%
\providecommand \url  [0]{\begingroup\@sanitize@url \@url }%
\providecommand \@url [1]{\endgroup\@href {#1}{\urlprefix }}%
\providecommand \urlprefix  [0]{URL }%
\providecommand \Eprint [0]{\href }%
\providecommand \doibase [0]{http://dx.doi.org/}%
\providecommand \selectlanguage [0]{\@gobble}%
\providecommand \bibinfo  [0]{\@secondoftwo}%
\providecommand \bibfield  [0]{\@secondoftwo}%
\providecommand \translation [1]{[#1]}%
\providecommand \BibitemOpen [0]{}%
\providecommand \bibitemStop [0]{}%
\providecommand \bibitemNoStop [0]{.\EOS\space}%
\providecommand \EOS [0]{\spacefactor3000\relax}%
\providecommand \BibitemShut  [1]{\csname bibitem#1\endcsname}%
\let\auto@bib@innerbib\@empty
\bibitem [{\citenamefont {Sutlive}\ \emph {et~al.}(2022)\citenamefont {Sutlive}, \citenamefont {Xiu}, \citenamefont {Chen}, \citenamefont {Gou}, \citenamefont {Xiong}, \citenamefont {Guo},\ and\ \citenamefont {Chen}}]{sutlive2022generation}%
  \BibitemOpen
  \bibfield  {author} {\bibinfo {author} {\bibfnamefont {J.}~\bibnamefont {Sutlive}}, \bibinfo {author} {\bibfnamefont {H.}~\bibnamefont {Xiu}}, \bibinfo {author} {\bibfnamefont {Y.}~\bibnamefont {Chen}}, \bibinfo {author} {\bibfnamefont {K.}~\bibnamefont {Gou}}, \bibinfo {author} {\bibfnamefont {F.}~\bibnamefont {Xiong}}, \bibinfo {author} {\bibfnamefont {M.}~\bibnamefont {Guo}}, \ and\ \bibinfo {author} {\bibfnamefont {Z.}~\bibnamefont {Chen}},\ }\href@noop {} {\bibfield  {journal} {\bibinfo  {journal} {Small}\ }\textbf {\bibinfo {volume} {18}},\ \bibinfo {pages} {2103466} (\bibinfo {year} {2022})}\BibitemShut {NoStop}%
\bibitem [{\citenamefont {Martin}\ and\ \citenamefont {Parkhurst}(2004)}]{martin2004parallels}%
  \BibitemOpen
  \bibfield  {author} {\bibinfo {author} {\bibfnamefont {P.}~\bibnamefont {Martin}}\ and\ \bibinfo {author} {\bibfnamefont {S.~M.}\ \bibnamefont {Parkhurst}},\ }\href {\doibase 10.1242/dev.01253} {\bibfield  {journal} {\bibinfo  {journal} {Development}\ }\textbf {\bibinfo {volume} {131}},\ \bibinfo {pages} {3021} (\bibinfo {year} {2004})}\BibitemShut {NoStop}%
\bibitem [{\citenamefont {Toyama}\ \emph {et~al.}(2008)\citenamefont {Toyama}, \citenamefont {Peralta}, \citenamefont {Wells}, \citenamefont {Kiehart},\ and\ \citenamefont {Edwards}}]{toyama2008apoptotic}%
  \BibitemOpen
  \bibfield  {author} {\bibinfo {author} {\bibfnamefont {Y.}~\bibnamefont {Toyama}}, \bibinfo {author} {\bibfnamefont {X.~G.}\ \bibnamefont {Peralta}}, \bibinfo {author} {\bibfnamefont {A.~R.}\ \bibnamefont {Wells}}, \bibinfo {author} {\bibfnamefont {D.~P.}\ \bibnamefont {Kiehart}}, \ and\ \bibinfo {author} {\bibfnamefont {G.~S.}\ \bibnamefont {Edwards}},\ }\href@noop {} {\bibfield  {journal} {\bibinfo  {journal} {Science}\ }\textbf {\bibinfo {volume} {321}},\ \bibinfo {pages} {1683} (\bibinfo {year} {2008})}\BibitemShut {NoStop}%
\bibitem [{\citenamefont {Jain}\ \emph {et~al.}(2014)\citenamefont {Jain}, \citenamefont {Martin},\ and\ \citenamefont {Stylianopoulos}}]{jain2014role}%
  \BibitemOpen
  \bibfield  {author} {\bibinfo {author} {\bibfnamefont {R.~K.}\ \bibnamefont {Jain}}, \bibinfo {author} {\bibfnamefont {J.~D.}\ \bibnamefont {Martin}}, \ and\ \bibinfo {author} {\bibfnamefont {T.}~\bibnamefont {Stylianopoulos}},\ }\href@noop {} {\bibfield  {journal} {\bibinfo  {journal} {Annu. Rev. Biomed. Eng.}\ }\textbf {\bibinfo {volume} {16}},\ \bibinfo {pages} {321} (\bibinfo {year} {2014})}\BibitemShut {NoStop}%
\bibitem [{\citenamefont {Wirtz}\ \emph {et~al.}(2011)\citenamefont {Wirtz}, \citenamefont {Konstantopoulos},\ and\ \citenamefont {Searson}}]{wirtz2011physics}%
  \BibitemOpen
  \bibfield  {author} {\bibinfo {author} {\bibfnamefont {D.}~\bibnamefont {Wirtz}}, \bibinfo {author} {\bibfnamefont {K.}~\bibnamefont {Konstantopoulos}}, \ and\ \bibinfo {author} {\bibfnamefont {P.~C.}\ \bibnamefont {Searson}},\ }\href@noop {} {\bibfield  {journal} {\bibinfo  {journal} {Nat. Rev. Cancer}\ }\textbf {\bibinfo {volume} {11}},\ \bibinfo {pages} {512} (\bibinfo {year} {2011})}\BibitemShut {NoStop}%
\bibitem [{\citenamefont {Trepat}\ \emph {et~al.}(2009{\natexlab{a}})\citenamefont {Trepat}, \citenamefont {Wasserman}, \citenamefont {Angelini}, \citenamefont {Millet}, \citenamefont {Weitz}, \citenamefont {Butler},\ and\ \citenamefont {Fredberg}}]{trepat2009physical}%
  \BibitemOpen
  \bibfield  {author} {\bibinfo {author} {\bibfnamefont {X.}~\bibnamefont {Trepat}}, \bibinfo {author} {\bibfnamefont {M.~R.}\ \bibnamefont {Wasserman}}, \bibinfo {author} {\bibfnamefont {T.~E.}\ \bibnamefont {Angelini}}, \bibinfo {author} {\bibfnamefont {E.}~\bibnamefont {Millet}}, \bibinfo {author} {\bibfnamefont {D.~A.}\ \bibnamefont {Weitz}}, \bibinfo {author} {\bibfnamefont {J.~P.}\ \bibnamefont {Butler}}, \ and\ \bibinfo {author} {\bibfnamefont {J.~J.}\ \bibnamefont {Fredberg}},\ }\href@noop {} {\bibfield  {journal} {\bibinfo  {journal} {Nat. Phys.}\ }\textbf {\bibinfo {volume} {5}},\ \bibinfo {pages} {426} (\bibinfo {year} {2009}{\natexlab{a}})}\BibitemShut {NoStop}%
\bibitem [{\citenamefont {Duclos}\ \emph {et~al.}(2017)\citenamefont {Duclos}, \citenamefont {Erlenk{\"a}mper}, \citenamefont {Joanny},\ and\ \citenamefont {Silberzan}}]{duclos2017topological}%
  \BibitemOpen
  \bibfield  {author} {\bibinfo {author} {\bibfnamefont {G.}~\bibnamefont {Duclos}}, \bibinfo {author} {\bibfnamefont {C.}~\bibnamefont {Erlenk{\"a}mper}}, \bibinfo {author} {\bibfnamefont {J.-F.}\ \bibnamefont {Joanny}}, \ and\ \bibinfo {author} {\bibfnamefont {P.}~\bibnamefont {Silberzan}},\ }\href@noop {} {\bibfield  {journal} {\bibinfo  {journal} {Nat. Phys.}\ }\textbf {\bibinfo {volume} {13}},\ \bibinfo {pages} {58} (\bibinfo {year} {2017})}\BibitemShut {NoStop}%
\bibitem [{\citenamefont {Balasubramaniam}\ \emph {et~al.}(2021)\citenamefont {Balasubramaniam}, \citenamefont {Doostmohammadi}, \citenamefont {Saw}, \citenamefont {Narayana}, \citenamefont {Mueller}, \citenamefont {Dang}, \citenamefont {Thomas}, \citenamefont {Gupta}, \citenamefont {Sonam}, \citenamefont {Yap} \emph {et~al.}}]{balasubramaniam2021investigating}%
  \BibitemOpen
  \bibfield  {author} {\bibinfo {author} {\bibfnamefont {L.}~\bibnamefont {Balasubramaniam}}, \bibinfo {author} {\bibfnamefont {A.}~\bibnamefont {Doostmohammadi}}, \bibinfo {author} {\bibfnamefont {T.~B.}\ \bibnamefont {Saw}}, \bibinfo {author} {\bibfnamefont {G.~H. N.~S.}\ \bibnamefont {Narayana}}, \bibinfo {author} {\bibfnamefont {R.}~\bibnamefont {Mueller}}, \bibinfo {author} {\bibfnamefont {T.}~\bibnamefont {Dang}}, \bibinfo {author} {\bibfnamefont {M.}~\bibnamefont {Thomas}}, \bibinfo {author} {\bibfnamefont {S.}~\bibnamefont {Gupta}}, \bibinfo {author} {\bibfnamefont {S.}~\bibnamefont {Sonam}}, \bibinfo {author} {\bibfnamefont {A.~S.}\ \bibnamefont {Yap}},  \emph {et~al.},\ }\href@noop {} {\bibfield  {journal} {\bibinfo  {journal} {Nat. Mater.}\ }\textbf {\bibinfo {volume} {20}},\ \bibinfo {pages} {1156} (\bibinfo {year} {2021})}\BibitemShut {NoStop}%
\bibitem [{\citenamefont {Saw}\ \emph {et~al.}(2018)\citenamefont {Saw}, \citenamefont {Xi}, \citenamefont {Ladoux},\ and\ \citenamefont {Lim}}]{saw2018biological}%
  \BibitemOpen
  \bibfield  {author} {\bibinfo {author} {\bibfnamefont {T.~B.}\ \bibnamefont {Saw}}, \bibinfo {author} {\bibfnamefont {W.}~\bibnamefont {Xi}}, \bibinfo {author} {\bibfnamefont {B.}~\bibnamefont {Ladoux}}, \ and\ \bibinfo {author} {\bibfnamefont {C.~T.}\ \bibnamefont {Lim}},\ }\href@noop {} {\bibfield  {journal} {\bibinfo  {journal} {Adv. Mater.}\ }\textbf {\bibinfo {volume} {30}},\ \bibinfo {pages} {1802579} (\bibinfo {year} {2018})}\BibitemShut {NoStop}%
\bibitem [{\citenamefont {Balasubramaniam}\ \emph {et~al.}(2022)\citenamefont {Balasubramaniam}, \citenamefont {M{\`e}ge},\ and\ \citenamefont {Ladoux}}]{balasubramaniam2022active}%
  \BibitemOpen
  \bibfield  {author} {\bibinfo {author} {\bibfnamefont {L.}~\bibnamefont {Balasubramaniam}}, \bibinfo {author} {\bibfnamefont {R.-M.}\ \bibnamefont {M{\`e}ge}}, \ and\ \bibinfo {author} {\bibfnamefont {B.}~\bibnamefont {Ladoux}},\ }\href@noop {} {\bibfield  {journal} {\bibinfo  {journal} {Curr. Opin. Genet. Dev.}\ }\textbf {\bibinfo {volume} {73}},\ \bibinfo {pages} {101897} (\bibinfo {year} {2022})}\BibitemShut {NoStop}%
\bibitem [{\citenamefont {Vafa}\ and\ \citenamefont {Mahadevan}(2022)}]{PhysRevLett.129.098102}%
  \BibitemOpen
  \bibfield  {author} {\bibinfo {author} {\bibfnamefont {F.}~\bibnamefont {Vafa}}\ and\ \bibinfo {author} {\bibfnamefont {L.}~\bibnamefont {Mahadevan}},\ }\href {\doibase 10.1103/PhysRevLett.129.098102} {\bibfield  {journal} {\bibinfo  {journal} {Phys. Rev. Lett.}\ }\textbf {\bibinfo {volume} {129}},\ \bibinfo {pages} {098102} (\bibinfo {year} {2022})}\BibitemShut {NoStop}%
\bibitem [{\citenamefont {Ramaswamy}(2017)}]{Ramaswamy_2017}%
  \BibitemOpen
  \bibfield  {author} {\bibinfo {author} {\bibfnamefont {S.}~\bibnamefont {Ramaswamy}},\ }\href {\doibase 10.1088/1742-5468/aa6bc5} {\bibfield  {journal} {\bibinfo  {journal} {J. Stat. Mech.: Theory Exp.}\ }\textbf {\bibinfo {volume} {2017}},\ \bibinfo {pages} {054002} (\bibinfo {year} {2017})}\BibitemShut {NoStop}%
\bibitem [{\citenamefont {Marchetti}\ \emph {et~al.}(2013)\citenamefont {Marchetti}, \citenamefont {Joanny}, \citenamefont {Ramaswamy}, \citenamefont {Liverpool}, \citenamefont {Prost}, \citenamefont {Rao},\ and\ \citenamefont {Simha}}]{RevModPhys.85.1143}%
  \BibitemOpen
  \bibfield  {author} {\bibinfo {author} {\bibfnamefont {M.~C.}\ \bibnamefont {Marchetti}}, \bibinfo {author} {\bibfnamefont {J.~F.}\ \bibnamefont {Joanny}}, \bibinfo {author} {\bibfnamefont {S.}~\bibnamefont {Ramaswamy}}, \bibinfo {author} {\bibfnamefont {T.~B.}\ \bibnamefont {Liverpool}}, \bibinfo {author} {\bibfnamefont {J.}~\bibnamefont {Prost}}, \bibinfo {author} {\bibfnamefont {M.}~\bibnamefont {Rao}}, \ and\ \bibinfo {author} {\bibfnamefont {R.~A.}\ \bibnamefont {Simha}},\ }\href {\doibase 10.1103/RevModPhys.85.1143} {\bibfield  {journal} {\bibinfo  {journal} {Rev. Mod. Phys.}\ }\textbf {\bibinfo {volume} {85}},\ \bibinfo {pages} {1143} (\bibinfo {year} {2013})}\BibitemShut {NoStop}%
\bibitem [{\citenamefont {Li}\ \emph {et~al.}(2017)\citenamefont {Li}, \citenamefont {Balagam}, \citenamefont {He}, \citenamefont {Lee}, \citenamefont {Igoshin},\ and\ \citenamefont {Levine}}]{li2017mechanism}%
  \BibitemOpen
  \bibfield  {author} {\bibinfo {author} {\bibfnamefont {X.}~\bibnamefont {Li}}, \bibinfo {author} {\bibfnamefont {R.}~\bibnamefont {Balagam}}, \bibinfo {author} {\bibfnamefont {T.-F.}\ \bibnamefont {He}}, \bibinfo {author} {\bibfnamefont {P.~P.}\ \bibnamefont {Lee}}, \bibinfo {author} {\bibfnamefont {O.~A.}\ \bibnamefont {Igoshin}}, \ and\ \bibinfo {author} {\bibfnamefont {H.}~\bibnamefont {Levine}},\ }\href@noop {} {\bibfield  {journal} {\bibinfo  {journal} {PNAS}\ }\textbf {\bibinfo {volume} {114}},\ \bibinfo {pages} {8974} (\bibinfo {year} {2017})}\BibitemShut {NoStop}%
\bibitem [{\citenamefont {Zhang}\ \emph {et~al.}(2021)\citenamefont {Zhang}, \citenamefont {Yang}, \citenamefont {Kreeger},\ and\ \citenamefont {Notbohm}}]{zhang2021}%
  \BibitemOpen
  \bibfield  {author} {\bibinfo {author} {\bibfnamefont {J.}~\bibnamefont {Zhang}}, \bibinfo {author} {\bibfnamefont {N.}~\bibnamefont {Yang}}, \bibinfo {author} {\bibfnamefont {P.~K.}\ \bibnamefont {Kreeger}}, \ and\ \bibinfo {author} {\bibfnamefont {J.}~\bibnamefont {Notbohm}},\ }\href@noop {} {\bibfield  {journal} {\bibinfo  {journal} {APL Bioeng.}\ }\textbf {\bibinfo {volume} {5}} (\bibinfo {year} {2021})}\BibitemShut {NoStop}%
\bibitem [{\citenamefont {Comelles}\ \emph {et~al.}(2021)\citenamefont {Comelles}, \citenamefont {Ss}, \citenamefont {Lu}, \citenamefont {Le~Maout}, \citenamefont {Anvitha}, \citenamefont {Salbreux}, \citenamefont {J{\"u}licher}, \citenamefont {Inamdar},\ and\ \citenamefont {Riveline}}]{comelles2021epithelial}%
  \BibitemOpen
  \bibfield  {author} {\bibinfo {author} {\bibfnamefont {J.}~\bibnamefont {Comelles}}, \bibinfo {author} {\bibfnamefont {S.}~\bibnamefont {Ss}}, \bibinfo {author} {\bibfnamefont {L.}~\bibnamefont {Lu}}, \bibinfo {author} {\bibfnamefont {E.}~\bibnamefont {Le~Maout}}, \bibinfo {author} {\bibfnamefont {S.}~\bibnamefont {Anvitha}}, \bibinfo {author} {\bibfnamefont {G.}~\bibnamefont {Salbreux}}, \bibinfo {author} {\bibfnamefont {F.}~\bibnamefont {J{\"u}licher}}, \bibinfo {author} {\bibfnamefont {M.~M.}\ \bibnamefont {Inamdar}}, \ and\ \bibinfo {author} {\bibfnamefont {D.}~\bibnamefont {Riveline}},\ }\href@noop {} {\bibfield  {journal} {\bibinfo  {journal} {eLife}\ }\textbf {\bibinfo {volume} {10}},\ \bibinfo {pages} {e57730} (\bibinfo {year} {2021})}\BibitemShut {NoStop}%
\bibitem [{\citenamefont {Saw}\ \emph {et~al.}(2017)\citenamefont {Saw}, \citenamefont {Doostmohammadi}, \citenamefont {Nier}, \citenamefont {Kocgozlu}, \citenamefont {Thampi}, \citenamefont {Toyama}, \citenamefont {Marcq}, \citenamefont {Lim}, \citenamefont {Yeomans},\ and\ \citenamefont {Ladoux}}]{saw2017topological}%
  \BibitemOpen
  \bibfield  {author} {\bibinfo {author} {\bibfnamefont {T.~B.}\ \bibnamefont {Saw}}, \bibinfo {author} {\bibfnamefont {A.}~\bibnamefont {Doostmohammadi}}, \bibinfo {author} {\bibfnamefont {V.}~\bibnamefont {Nier}}, \bibinfo {author} {\bibfnamefont {L.}~\bibnamefont {Kocgozlu}}, \bibinfo {author} {\bibfnamefont {S.}~\bibnamefont {Thampi}}, \bibinfo {author} {\bibfnamefont {Y.}~\bibnamefont {Toyama}}, \bibinfo {author} {\bibfnamefont {P.}~\bibnamefont {Marcq}}, \bibinfo {author} {\bibfnamefont {C.~T.}\ \bibnamefont {Lim}}, \bibinfo {author} {\bibfnamefont {J.~M.}\ \bibnamefont {Yeomans}}, \ and\ \bibinfo {author} {\bibfnamefont {B.}~\bibnamefont {Ladoux}},\ }\href@noop {} {\bibfield  {journal} {\bibinfo  {journal} {Nature}\ }\textbf {\bibinfo {volume} {544}},\ \bibinfo {pages} {212} (\bibinfo {year} {2017})}\BibitemShut {NoStop}%
\bibitem [{\citenamefont {Ascione}\ \emph {et~al.}(2023)\citenamefont {Ascione}, \citenamefont {Caserta}, \citenamefont {Esposito}, \citenamefont {Villella}, \citenamefont {Maiuri}, \citenamefont {Nejad}, \citenamefont {Doostmohammadi}, \citenamefont {Yeomans},\ and\ \citenamefont {Guido}}]{ascione2022collective}%
  \BibitemOpen
  \bibfield  {author} {\bibinfo {author} {\bibfnamefont {F.}~\bibnamefont {Ascione}}, \bibinfo {author} {\bibfnamefont {S.}~\bibnamefont {Caserta}}, \bibinfo {author} {\bibfnamefont {S.}~\bibnamefont {Esposito}}, \bibinfo {author} {\bibfnamefont {V.~R.}\ \bibnamefont {Villella}}, \bibinfo {author} {\bibfnamefont {L.}~\bibnamefont {Maiuri}}, \bibinfo {author} {\bibfnamefont {M.~R.}\ \bibnamefont {Nejad}}, \bibinfo {author} {\bibfnamefont {A.}~\bibnamefont {Doostmohammadi}}, \bibinfo {author} {\bibfnamefont {J.~M.}\ \bibnamefont {Yeomans}}, \ and\ \bibinfo {author} {\bibfnamefont {S.}~\bibnamefont {Guido}},\ }\href@noop {} {\bibfield  {journal} {\bibinfo  {journal} {J. R. Soc. Interface}\ }\textbf {\bibinfo {volume} {20}},\ \bibinfo {pages} {20220719} (\bibinfo {year} {2023})}\BibitemShut {NoStop}%
\bibitem [{\citenamefont {Blanch-Mercader}\ \emph {et~al.}(2018)\citenamefont {Blanch-Mercader}, \citenamefont {Yashunsky}, \citenamefont {Garcia}, \citenamefont {Duclos}, \citenamefont {Giomi},\ and\ \citenamefont {Silberzan}}]{PhysRevLett.120.208101}%
  \BibitemOpen
  \bibfield  {author} {\bibinfo {author} {\bibfnamefont {C.}~\bibnamefont {Blanch-Mercader}}, \bibinfo {author} {\bibfnamefont {V.}~\bibnamefont {Yashunsky}}, \bibinfo {author} {\bibfnamefont {S.}~\bibnamefont {Garcia}}, \bibinfo {author} {\bibfnamefont {G.}~\bibnamefont {Duclos}}, \bibinfo {author} {\bibfnamefont {L.}~\bibnamefont {Giomi}}, \ and\ \bibinfo {author} {\bibfnamefont {P.}~\bibnamefont {Silberzan}},\ }\href {\doibase 10.1103/PhysRevLett.120.208101} {\bibfield  {journal} {\bibinfo  {journal} {Phys. Rev. Lett.}\ }\textbf {\bibinfo {volume} {120}},\ \bibinfo {pages} {208101} (\bibinfo {year} {2018})}\BibitemShut {NoStop}%
\bibitem [{\citenamefont {Duclos}\ \emph {et~al.}(2018)\citenamefont {Duclos}, \citenamefont {Blanch-Mercader}, \citenamefont {Yashunsky}, \citenamefont {Salbreux}, \citenamefont {Joanny}, \citenamefont {Prost},\ and\ \citenamefont {Silberzan}}]{duclos2018spontaneous}%
  \BibitemOpen
  \bibfield  {author} {\bibinfo {author} {\bibfnamefont {G.}~\bibnamefont {Duclos}}, \bibinfo {author} {\bibfnamefont {C.}~\bibnamefont {Blanch-Mercader}}, \bibinfo {author} {\bibfnamefont {V.}~\bibnamefont {Yashunsky}}, \bibinfo {author} {\bibfnamefont {G.}~\bibnamefont {Salbreux}}, \bibinfo {author} {\bibfnamefont {J.-F.}\ \bibnamefont {Joanny}}, \bibinfo {author} {\bibfnamefont {J.}~\bibnamefont {Prost}}, \ and\ \bibinfo {author} {\bibfnamefont {P.}~\bibnamefont {Silberzan}},\ }\href@noop {} {\bibfield  {journal} {\bibinfo  {journal} {Nat. Phys.}\ }\textbf {\bibinfo {volume} {14}},\ \bibinfo {pages} {728} (\bibinfo {year} {2018})}\BibitemShut {NoStop}%
\bibitem [{\citenamefont {Simha}\ and\ \citenamefont {Ramaswamy}(2002)}]{Simha02}%
  \BibitemOpen
  \bibfield  {author} {\bibinfo {author} {\bibfnamefont {R.~A.}\ \bibnamefont {Simha}}\ and\ \bibinfo {author} {\bibfnamefont {S.}~\bibnamefont {Ramaswamy}},\ }\href {\doibase 10.1103/PhysRevLett.89.058101} {\bibfield  {journal} {\bibinfo  {journal} {Phys. Rev. Lett.}\ }\textbf {\bibinfo {volume} {89}},\ \bibinfo {pages} {058101} (\bibinfo {year} {2002})}\BibitemShut {NoStop}%
\bibitem [{\citenamefont {Doostmohammadi}\ \emph {et~al.}(2020)\citenamefont {Doostmohammadi}, \citenamefont {Ign{\'e}s-Mullol}, \citenamefont {Yeomans},\ and\ \citenamefont {Sagu{\'e}s}}]{Doostmohammadi18}%
  \BibitemOpen
  \bibfield  {author} {\bibinfo {author} {\bibfnamefont {A.}~\bibnamefont {Doostmohammadi}}, \bibinfo {author} {\bibfnamefont {J.}~\bibnamefont {Ign{\'e}s-Mullol}}, \bibinfo {author} {\bibfnamefont {J.~M.}\ \bibnamefont {Yeomans}}, \ and\ \bibinfo {author} {\bibfnamefont {F.}~\bibnamefont {Sagu{\'e}s}},\ }\href@noop {} {\bibfield  {journal} {\bibinfo  {journal} {Nature Comms.}\ }\textbf {\bibinfo {volume} {9}},\ \bibinfo {pages} {3246} (\bibinfo {year} {2020})}\BibitemShut {NoStop}%
\bibitem [{\citenamefont {Yashunsky}\ \emph {et~al.}(2022)\citenamefont {Yashunsky}, \citenamefont {Pearce}, \citenamefont {Blanch-Mercader}, \citenamefont {Ascione}, \citenamefont {Silberzan},\ and\ \citenamefont {Giomi}}]{PhysRevX.12.041017}%
  \BibitemOpen
  \bibfield  {author} {\bibinfo {author} {\bibfnamefont {V.}~\bibnamefont {Yashunsky}}, \bibinfo {author} {\bibfnamefont {D.~J.~G.}\ \bibnamefont {Pearce}}, \bibinfo {author} {\bibfnamefont {C.}~\bibnamefont {Blanch-Mercader}}, \bibinfo {author} {\bibfnamefont {F.}~\bibnamefont {Ascione}}, \bibinfo {author} {\bibfnamefont {P.}~\bibnamefont {Silberzan}}, \ and\ \bibinfo {author} {\bibfnamefont {L.}~\bibnamefont {Giomi}},\ }\href {\doibase 10.1103/PhysRevX.12.041017} {\bibfield  {journal} {\bibinfo  {journal} {Phys. Rev. X}\ }\textbf {\bibinfo {volume} {12}},\ \bibinfo {pages} {041017} (\bibinfo {year} {2022})}\BibitemShut {NoStop}%
\bibitem [{\citenamefont {de~Gennes}\ and\ \citenamefont {Prost}(1995)}]{DeGennesBook}%
  \BibitemOpen
  \bibfield  {author} {\bibinfo {author} {\bibfnamefont {P.~G.}\ \bibnamefont {de~Gennes}}\ and\ \bibinfo {author} {\bibfnamefont {J.}~\bibnamefont {Prost}},\ }\href@noop {} {\emph {\bibinfo {title} {The Physics of Liquid Crystals}}}\ (\bibinfo  {publisher} {Oxford University Press},\ \bibinfo {year} {1995})\BibitemShut {NoStop}%
\bibitem [{\citenamefont {Pellegrin}\ and\ \citenamefont {Mellor}(2007)}]{Pellegrin07}%
  \BibitemOpen
  \bibfield  {author} {\bibinfo {author} {\bibfnamefont {S.}~\bibnamefont {Pellegrin}}\ and\ \bibinfo {author} {\bibfnamefont {H.}~\bibnamefont {Mellor}},\ }\href {\doibase 10.1242/jcs.018473} {\bibfield  {journal} {\bibinfo  {journal} {J. Cell Sci.}\ }\textbf {\bibinfo {volume} {120}},\ \bibinfo {pages} {3491} (\bibinfo {year} {2007})}\BibitemShut {NoStop}%
\bibitem [{\citenamefont {Saraswathibhatla}\ and\ \citenamefont {Notbohm}(2020)}]{saraswathibhatla2020}%
  \BibitemOpen
  \bibfield  {author} {\bibinfo {author} {\bibfnamefont {A.}~\bibnamefont {Saraswathibhatla}}\ and\ \bibinfo {author} {\bibfnamefont {J.}~\bibnamefont {Notbohm}},\ }\href@noop {} {\bibfield  {journal} {\bibinfo  {journal} {Phys. Rev. X}\ }\textbf {\bibinfo {volume} {10}},\ \bibinfo {pages} {011016} (\bibinfo {year} {2020})}\BibitemShut {NoStop}%
\bibitem [{\citenamefont {Saraswathibhatla}\ \emph {et~al.}(2020)\citenamefont {Saraswathibhatla}, \citenamefont {Galles},\ and\ \citenamefont {Notbohm}}]{saraswathibhatla2020scidata}%
  \BibitemOpen
  \bibfield  {author} {\bibinfo {author} {\bibfnamefont {A.}~\bibnamefont {Saraswathibhatla}}, \bibinfo {author} {\bibfnamefont {E.}~\bibnamefont {Galles}}, \ and\ \bibinfo {author} {\bibfnamefont {J.}~\bibnamefont {Notbohm}},\ }\href@noop {} {\bibfield  {journal} {\bibinfo  {journal} {Sci. Data}\ }\textbf {\bibinfo {volume} {7}},\ \bibinfo {pages} {197} (\bibinfo {year} {2020})}\BibitemShut {NoStop}%
\bibitem [{\citenamefont {Dembo}\ and\ \citenamefont {Wang}(2011)}]{dembo1999}%
  \BibitemOpen
  \bibfield  {author} {\bibinfo {author} {\bibfnamefont {M.}~\bibnamefont {Dembo}}\ and\ \bibinfo {author} {\bibfnamefont {Y.-L.}\ \bibnamefont {Wang}},\ }\href@noop {} {\bibfield  {journal} {\bibinfo  {journal} {Biophys. J.}\ }\textbf {\bibinfo {volume} {76}},\ \bibinfo {pages} {2307} (\bibinfo {year} {2011})}\BibitemShut {NoStop}%
\bibitem [{\citenamefont {Bar-Kochba}\ \emph {et~al.}(2015)\citenamefont {Bar-Kochba}, \citenamefont {Toyjanova}, \citenamefont {Andrews}, \citenamefont {Kim},\ and\ \citenamefont {Franck}}]{Bar-Kochba2015}%
  \BibitemOpen
  \bibfield  {author} {\bibinfo {author} {\bibfnamefont {E.}~\bibnamefont {Bar-Kochba}}, \bibinfo {author} {\bibfnamefont {J.}~\bibnamefont {Toyjanova}}, \bibinfo {author} {\bibfnamefont {E.}~\bibnamefont {Andrews}}, \bibinfo {author} {\bibfnamefont {K.-S.}\ \bibnamefont {Kim}}, \ and\ \bibinfo {author} {\bibfnamefont {C.}~\bibnamefont {Franck}},\ }\href@noop {} {\bibfield  {journal} {\bibinfo  {journal} {Exp. Mech.}\ }\textbf {\bibinfo {volume} {55}},\ \bibinfo {pages} {261} (\bibinfo {year} {2015})}\BibitemShut {NoStop}%
\bibitem [{\citenamefont {Butler}\ \emph {et~al.}(2002)\citenamefont {Butler}, \citenamefont {Toli{\'c}-N{\o}rrelykke}, \citenamefont {Fabry},\ and\ \citenamefont {Fredberg}}]{butler2002}%
  \BibitemOpen
  \bibfield  {author} {\bibinfo {author} {\bibfnamefont {J.}~\bibnamefont {Butler}}, \bibinfo {author} {\bibfnamefont {I.}~\bibnamefont {Toli{\'c}-N{\o}rrelykke}}, \bibinfo {author} {\bibfnamefont {B.}~\bibnamefont {Fabry}}, \ and\ \bibinfo {author} {\bibfnamefont {J.}~\bibnamefont {Fredberg}},\ }\href@noop {} {\bibfield  {journal} {\bibinfo  {journal} {Am. J. Physiol. Cell Physiol.}\ }\textbf {\bibinfo {volume} {282}},\ \bibinfo {pages} {C595} (\bibinfo {year} {2002})}\BibitemShut {NoStop}%
\bibitem [{\citenamefont {Del~Alamo}\ \emph {et~al.}(2007)\citenamefont {Del~Alamo}, \citenamefont {Meili}, \citenamefont {Alonso-Latorre}, \citenamefont {Rodr{\'\i}guez-Rodr{\'\i}guez}, \citenamefont {Aliseda}, \citenamefont {Firteland},\ and\ \citenamefont {Lasheras}}]{delalamo2007}%
  \BibitemOpen
  \bibfield  {author} {\bibinfo {author} {\bibfnamefont {J.~C.}\ \bibnamefont {Del~Alamo}}, \bibinfo {author} {\bibfnamefont {R.}~\bibnamefont {Meili}}, \bibinfo {author} {\bibfnamefont {B.}~\bibnamefont {Alonso-Latorre}}, \bibinfo {author} {\bibfnamefont {J.}~\bibnamefont {Rodr{\'\i}guez-Rodr{\'\i}guez}}, \bibinfo {author} {\bibfnamefont {A.}~\bibnamefont {Aliseda}}, \bibinfo {author} {\bibfnamefont {R.~A.}\ \bibnamefont {Firteland}}, \ and\ \bibinfo {author} {\bibfnamefont {J.~C.}\ \bibnamefont {Lasheras}},\ }\href@noop {} {\bibfield  {journal} {\bibinfo  {journal} {PNAS}\ }\textbf {\bibinfo {volume} {104}},\ \bibinfo {pages} {13343} (\bibinfo {year} {2007})}\BibitemShut {NoStop}%
\bibitem [{\citenamefont {Trepat}\ \emph {et~al.}(2009{\natexlab{b}})\citenamefont {Trepat}, \citenamefont {Wasserman}, \citenamefont {Angelini}, \citenamefont {Millet}, \citenamefont {Weitz}, \citenamefont {Butler},\ and\ \citenamefont {Fredberg}}]{trepat2009}%
  \BibitemOpen
  \bibfield  {author} {\bibinfo {author} {\bibfnamefont {X.}~\bibnamefont {Trepat}}, \bibinfo {author} {\bibfnamefont {M.}~\bibnamefont {Wasserman}}, \bibinfo {author} {\bibfnamefont {T.}~\bibnamefont {Angelini}}, \bibinfo {author} {\bibfnamefont {E.}~\bibnamefont {Millet}}, \bibinfo {author} {\bibfnamefont {D.}~\bibnamefont {Weitz}}, \bibinfo {author} {\bibfnamefont {J.}~\bibnamefont {Butler}}, \ and\ \bibinfo {author} {\bibfnamefont {J.}~\bibnamefont {Fredberg}},\ }\href@noop {} {\bibfield  {journal} {\bibinfo  {journal} {Nat. Phys.}\ }\textbf {\bibinfo {volume} {5}},\ \bibinfo {pages} {426} (\bibinfo {year} {2009}{\natexlab{b}})}\BibitemShut {NoStop}%
\bibitem [{\citenamefont {Tambe}\ \emph {et~al.}(2011)\citenamefont {Tambe}, \citenamefont {Corey~Hardin}, \citenamefont {Angelini}, \citenamefont {Rajendran}, \citenamefont {Park}, \citenamefont {Serra-Picamal}, \citenamefont {Zhou}, \citenamefont {Zaman}, \citenamefont {Butler}, \citenamefont {Weitz},\ and\ \citenamefont {Fredberg}}]{tambe2011}%
  \BibitemOpen
  \bibfield  {author} {\bibinfo {author} {\bibfnamefont {D.}~\bibnamefont {Tambe}}, \bibinfo {author} {\bibfnamefont {C.}~\bibnamefont {Corey~Hardin}}, \bibinfo {author} {\bibfnamefont {T.}~\bibnamefont {Angelini}}, \bibinfo {author} {\bibfnamefont {K.}~\bibnamefont {Rajendran}}, \bibinfo {author} {\bibfnamefont {C.}~\bibnamefont {Park}}, \bibinfo {author} {\bibfnamefont {X.}~\bibnamefont {Serra-Picamal}}, \bibinfo {author} {\bibfnamefont {E.}~\bibnamefont {Zhou}}, \bibinfo {author} {\bibfnamefont {M.}~\bibnamefont {Zaman}}, \bibinfo {author} {\bibfnamefont {J.}~\bibnamefont {Butler}}, \bibinfo {author} {\bibfnamefont {D.}~\bibnamefont {Weitz}}, \ and\ \bibinfo {author} {\bibfnamefont {J.}~\bibnamefont {Fredberg}},\ }\href@noop {} {\bibfield  {journal} {\bibinfo  {journal} {Nat. Mater}\ }\textbf {\bibinfo {volume} {6}},\ \bibinfo {pages} {469} (\bibinfo {year} {2011})}\BibitemShut {NoStop}%
\bibitem [{\citenamefont {Tambe}\ \emph {et~al.}(2013)\citenamefont {Tambe}, \citenamefont {Croutelle}, \citenamefont {Trepat}, \citenamefont {Park}, \citenamefont {Kim}, \citenamefont {Millet}, \citenamefont {Butler},\ and\ \citenamefont {Fredberg}}]{tambe2013}%
  \BibitemOpen
  \bibfield  {author} {\bibinfo {author} {\bibfnamefont {D.~T.}\ \bibnamefont {Tambe}}, \bibinfo {author} {\bibfnamefont {U.}~\bibnamefont {Croutelle}}, \bibinfo {author} {\bibfnamefont {X.}~\bibnamefont {Trepat}}, \bibinfo {author} {\bibfnamefont {C.~Y.}\ \bibnamefont {Park}}, \bibinfo {author} {\bibfnamefont {J.~H.}\ \bibnamefont {Kim}}, \bibinfo {author} {\bibfnamefont {E.}~\bibnamefont {Millet}}, \bibinfo {author} {\bibfnamefont {J.~P.}\ \bibnamefont {Butler}}, \ and\ \bibinfo {author} {\bibfnamefont {J.~J.}\ \bibnamefont {Fredberg}},\ }\href@noop {} {\bibfield  {journal} {\bibinfo  {journal} {Plos One}\ }\textbf {\bibinfo {volume} {8}},\ \bibinfo {pages} {e55172} (\bibinfo {year} {2013})}\BibitemShut {NoStop}%
\bibitem [{\citenamefont {Vromans}\ and\ \citenamefont {Giomi}(2016)}]{vromans2016orientational}%
  \BibitemOpen
  \bibfield  {author} {\bibinfo {author} {\bibfnamefont {A.~J.}\ \bibnamefont {Vromans}}\ and\ \bibinfo {author} {\bibfnamefont {L.}~\bibnamefont {Giomi}},\ }\href@noop {} {\bibfield  {journal} {\bibinfo  {journal} {Soft Matter}\ }\textbf {\bibinfo {volume} {12}},\ \bibinfo {pages} {6490} (\bibinfo {year} {2016})}\BibitemShut {NoStop}%
\bibitem [{\citenamefont {Mertz}\ \emph {et~al.}(2012)\citenamefont {Mertz}, \citenamefont {Banerjee}, \citenamefont {Che}, \citenamefont {German}, \citenamefont {Xu}, \citenamefont {Hyland}, \citenamefont {Marchetti}, \citenamefont {Horsley}, \citenamefont {Dufresne} \emph {et~al.}}]{mertz2012}%
  \BibitemOpen
  \bibfield  {author} {\bibinfo {author} {\bibfnamefont {A.~F.}\ \bibnamefont {Mertz}}, \bibinfo {author} {\bibfnamefont {S.}~\bibnamefont {Banerjee}}, \bibinfo {author} {\bibfnamefont {Y.}~\bibnamefont {Che}}, \bibinfo {author} {\bibfnamefont {G.~K.}\ \bibnamefont {German}}, \bibinfo {author} {\bibfnamefont {Y.}~\bibnamefont {Xu}}, \bibinfo {author} {\bibfnamefont {C.}~\bibnamefont {Hyland}}, \bibinfo {author} {\bibfnamefont {M.~C.}\ \bibnamefont {Marchetti}}, \bibinfo {author} {\bibfnamefont {V.}~\bibnamefont {Horsley}}, \bibinfo {author} {\bibfnamefont {E.~R.}\ \bibnamefont {Dufresne}},  \emph {et~al.},\ }\href@noop {} {\bibfield  {journal} {\bibinfo  {journal} {Phys. Rev. Lett.}\ }\textbf {\bibinfo {volume} {108}},\ \bibinfo {pages} {198101} (\bibinfo {year} {2012})}\BibitemShut {NoStop}%
\bibitem [{\citenamefont {Saraswathibhatla}\ \emph {et~al.}(2021)\citenamefont {Saraswathibhatla}, \citenamefont {Henkes}, \citenamefont {Galles}, \citenamefont {Sknepnek},\ and\ \citenamefont {Notbohm}}]{saraswathibhatla2021}%
  \BibitemOpen
  \bibfield  {author} {\bibinfo {author} {\bibfnamefont {A.}~\bibnamefont {Saraswathibhatla}}, \bibinfo {author} {\bibfnamefont {S.}~\bibnamefont {Henkes}}, \bibinfo {author} {\bibfnamefont {E.~E.}\ \bibnamefont {Galles}}, \bibinfo {author} {\bibfnamefont {R.}~\bibnamefont {Sknepnek}}, \ and\ \bibinfo {author} {\bibfnamefont {J.}~\bibnamefont {Notbohm}},\ }\href@noop {} {\bibfield  {journal} {\bibinfo  {journal} {Extreme Mech. Lett.}\ }\textbf {\bibinfo {volume} {48}},\ \bibinfo {pages} {101438} (\bibinfo {year} {2021})}\BibitemShut {NoStop}%
\bibitem [{\citenamefont {Notbohm}\ \emph {et~al.}(2016)\citenamefont {Notbohm}, \citenamefont {Banerjee}, \citenamefont {Utuje}, \citenamefont {Gweon}, \citenamefont {Jang}, \citenamefont {Park}, \citenamefont {Shin}, \citenamefont {Butler}, \citenamefont {Fredberg},\ and\ \citenamefont {Marchetti}}]{notbohm2016}%
  \BibitemOpen
  \bibfield  {author} {\bibinfo {author} {\bibfnamefont {J.}~\bibnamefont {Notbohm}}, \bibinfo {author} {\bibfnamefont {S.}~\bibnamefont {Banerjee}}, \bibinfo {author} {\bibfnamefont {K.~J.}\ \bibnamefont {Utuje}}, \bibinfo {author} {\bibfnamefont {B.}~\bibnamefont {Gweon}}, \bibinfo {author} {\bibfnamefont {H.}~\bibnamefont {Jang}}, \bibinfo {author} {\bibfnamefont {Y.}~\bibnamefont {Park}}, \bibinfo {author} {\bibfnamefont {J.}~\bibnamefont {Shin}}, \bibinfo {author} {\bibfnamefont {J.~P.}\ \bibnamefont {Butler}}, \bibinfo {author} {\bibfnamefont {J.~J.}\ \bibnamefont {Fredberg}}, \ and\ \bibinfo {author} {\bibfnamefont {M.~C.}\ \bibnamefont {Marchetti}},\ }\href@noop {} {\bibfield  {journal} {\bibinfo  {journal} {Biophys. J.}\ }\textbf {\bibinfo {volume} {110}},\ \bibinfo {pages} {2729} (\bibinfo {year} {2016})}\BibitemShut {NoStop}%
\bibitem [{\citenamefont {Beris}\ and\ \citenamefont {Edwards}(1994)}]{beris1994thermodynamics}%
  \BibitemOpen
  \bibfield  {author} {\bibinfo {author} {\bibfnamefont {A.~N.}\ \bibnamefont {Beris}}\ and\ \bibinfo {author} {\bibfnamefont {B.~J.}\ \bibnamefont {Edwards}},\ }\href@noop {} {\emph {\bibinfo {title} {Thermodynamics of Flowing Systems: With Internal Microstructure}}}\ (\bibinfo  {publisher} {Oxford University Press},\ \bibinfo {year} {1994})\BibitemShut {NoStop}%
\bibitem [{\citenamefont {Marenduzzo}\ \emph {et~al.}(2007)\citenamefont {Marenduzzo}, \citenamefont {Orlandini}, \citenamefont {Cates},\ and\ \citenamefont {Yeomans}}]{Marenduzzo2007}%
  \BibitemOpen
  \bibfield  {author} {\bibinfo {author} {\bibfnamefont {D.}~\bibnamefont {Marenduzzo}}, \bibinfo {author} {\bibfnamefont {E.}~\bibnamefont {Orlandini}}, \bibinfo {author} {\bibfnamefont {M.~E.}\ \bibnamefont {Cates}}, \ and\ \bibinfo {author} {\bibfnamefont {J.~M.}\ \bibnamefont {Yeomans}},\ }\href@noop {} {\bibfield  {journal} {\bibinfo  {journal} {Phys. Rev. E}\ }\textbf {\bibinfo {volume} {76}},\ \bibinfo {pages} {031921} (\bibinfo {year} {2007})}\BibitemShut {NoStop}%
\end{thebibliography}%

\begin{figure*}[h] 
    \centering
     \includegraphics[width=0.99\textwidth]{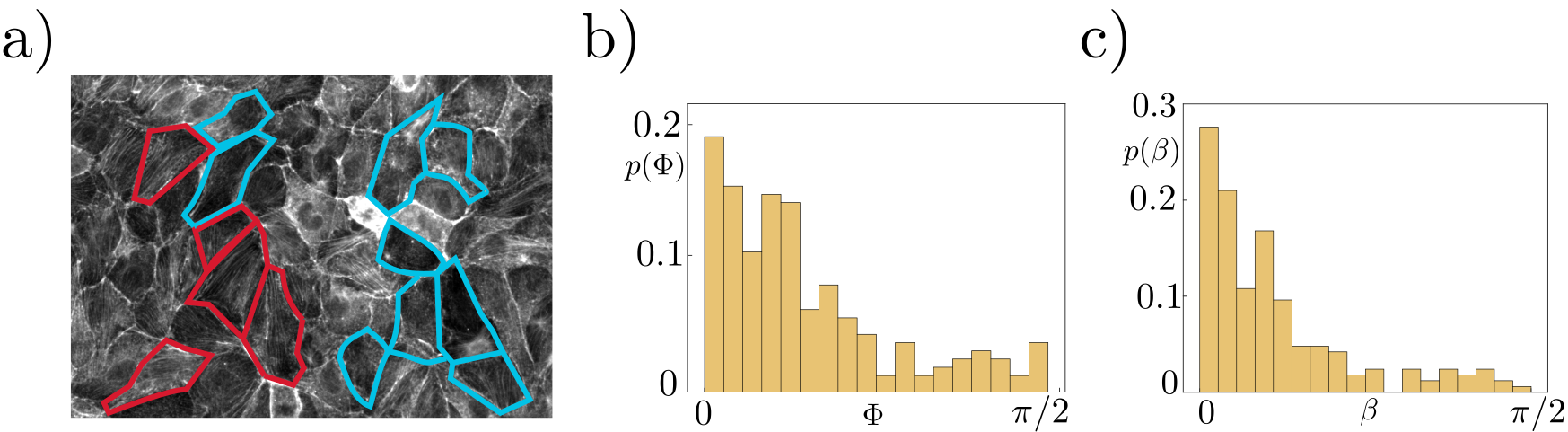}
    \caption{(a) Enlarged snapshot of the tissue in Fig. 1 where blue (red) contours specify extensile (contractile) cells. In general, the actin is clearer in the contractile cells, where shape, the principal axis of contractile stress, and the actin tend to be aligned. (b) Probability distribution of the angle between actin and the principal axis of contractile stress $\Phi$ in cells where the actin was clear enough to enable a measurement. (c) Probability distribution of the angle between actin and the direction of cell shape elongation $\beta$ in cells where the actin was clear enough to enable a measurement.}\label{fibers}
\end{figure*}

\newpage
\noindent
{\bf Supplementary Material:}\\
~\\
\noindent
Using four independent MDCK samples we fixed cells for fluorescent imaging of actin fibres.
Visual observations showed that it was rarely possible to visually ascertain an unambiguous direction of the stress fibres in cells with large misalignment angle $\theta$, whereas stress fibres were much clearer in cases where they were aligned along the long axis of the cell. This is illustrated in Fig.~\ref{fibers}.

\end{document}